\documentclass[a4paper]{article}
\makeatletter
\def\@seccntformat#1{\@ifundefined{#1@cntformat}%
   {\csname the#1\endcsname\quad}  % default
   {\csname #1@cntformat\endcsname}% enable individual control
}
\let\oldappendix\appendix %% save current definition of \appendix
\renewcommand\appendix{%
    \oldappendix
    \newcommand{\section@cntformat}{\appendixname~\thesection\quad}
}
\makeatother

\usepackage[dvipdfmx]{graphicx}

\usepackage{cite}
\usepackage[normalem]{ulem}
\usepackage{comment}

\usepackage[usenames,dvipsnames]{xcolor} % extra colors

\usepackage{amssymb,amsfonts,amsmath}
\usepackage[top=25truemm,bottom=25truemm,left=20truemm,right=20truemm]{geometry}
\usepackage{bm}
\usepackage{cases}
\usepackage[mathlines]{lineno}
\usepackage{setspace}
\title{Conditions for the existence of zero-determinant strategies\\ under observation errors in repeated games}
\bigskip
\author{Azumi Mamiya${}^{1}$, Daiki Miyagawa${}^{2}$ and Genki Ichinose${}^{2*}$
\ \\
\ \\
${}^{1}$
Nagoya Works, Mitsubishi Electric Corporation, \\5-1-14, Yada-minami, Higashi-ku, Nagoya,  461-8670, Japan\\
${}^{2}$
Department of Mathematical and Systems Engineering, Shizuoka University, \\3-5-1 Johoku, Naka-ku, Hamamatsu, 432-8561, Japan\\
$^*$ Corresponding author (ichinose.genki@shizuoka.ac.jp)}

\begin{document}

\maketitle

\section*{Abstract}

Repeated games are useful models to analyze long term interactions of living species and complex social phenomena. 
Zero-determinant (ZD) strategies in repeated games discovered by Press and Dyson in 2012 enforce a linear payoff relationship between a focal player and the opponent.
This linear relationship can be set arbitrarily by a ZD player. Hence, a subclass of ZD strategies can fix the opponent's expected payoff and another subclass of the strategies can exceed the opponent for the expected payoff.
Since this discovery, theories for ZD strategies are extended to cope with various natural situations.
It is especially important to consider the theory of ZD strategies for repeated games with a discount factor and observation errors because it allows the theory to be applicable in the real world.
Recent studies revealed their existence of ZD strategies even in repeated games with both factors.
However, the conditions for the existence has not been sufficiently analyzed.
Here, we mathematically analyzed the conditions in repeated games with both factors.
First, we derived the thresholds of a discount factor and observation errors which ensure the existence of Equalizer and positively correlated ZD (pcZD) strategies, which are well-known subclasses of ZD strategies.
We found that ZD strategies exist only when a discount factor remains high as the error rates increase.
Next, we derived the conditions for the expected payoff of the opponent enforced by Equalizer as well as the conditions for the slope and base line payoff of linear lines enforced by pcZD.
As a result, we found that, as error rates increase or a discount factor decreases, the conditions for the linear line that Equalizer or pcZD can enforce become strict.

\section*{Keywords}
Prisoner's dilemma, repeated games, discount factor, observation errors, zero-determinant strategies

\section{Introduction\label{sec:introduction}}

%In 2012, in Repeated Prisoner's Dilemma (RPD) games with perfect monitoring, Press and Dyson discovered the existence of Zero-determinant (ZD) strategies. Such strategies  unilaterally enforce a linear constraint on the payoffs of two players. If the determinant of a particular matrix becomes zero, a linear constraint on the payoffs is realized. ZD strategies are named such the curious name because they are derived by using the above condition. The shape of this constraint is unilaterally under a focal player's control.

%Repeated games with imperfect private monitoring is the modern theoretical framework to analyze interactions of living species and complex social phenomena. In imperfect monitoring, it is assumed that each player observes signals depend on previous the opponent's  action instead of being unable to directly observe the action. Moreover, in private monitoring, each player receives private signals about the opponent's actions.

%ZD strategies have Extortion, Generous, Equalizer as the subclass strategies.

Mutual cooperation, competition, and exploitation are often observed in human societies and interactions of biological organisms.
Cooperation gives benefits to others while exploitation (defection) obtains the benefits without paying costs.
The Repeated Prisoner's Dilemma (RPD) game is a theoretical framework to study strategic behaviors in social dilemmas \cite{MailathSamuelson2006book}.
The rule is simple: First, each player selects Cooperation (C) or Defection (D), respectively. Then, depending on the combination, the payoff is allocated to each player.
Mutual cooperation yields higher benefits than mutual defection. However, from the individual point of view, cooperation is always exploited by defection.
Thus, in the one-shot prisoner's dilemma, defection is the unique Nash equilibrium, resulting in mutual defection between self-interested players.
The situation changes when the game is repeated.
In the RPD, cooperation becomes a possible choice because future payoffs are important.
This is called direct reciprocity \cite{Trivers1971QRevBiol,Nowak2006book,Sigmund2010book}.
Cooperative behaviors in the RPD games have been studied in game theory and evolutionary games.

In 2012, the discovery of Zero-determinant (ZD) strategies by Press and Dyson changed the way of studying direct reciprocity \cite{Press2012PNAS}.
In contrast to the approach used in evolutionary games where a population of individuals is considered, they studied a one-to-one interaction in the RPD and showed that one of the two players can unilaterally enforce a linear relationship to the opponent.
Recently, the general properties of ZD strategies and broader classes of strategies including ZD have been studied and these properties have gradually been unraveled \cite{Hilbe2018NatHumBehav, Murase2018JTheorBiol, Murase2020SciRep, ChenZinger2014JTheorBiol, UedaTanaka2020PlosOne, Ueda2021JPhysSocJpn}.

Due to the linear payoff relationship, a subclass of ZD called Equalizer can fix the opponent's payoff. Another subclass of ZD called Extortion can obtain a larger or at least equal payoff than the opponent. The slope of the line between Extortion and the opponent is positive, which means that if the opponent adapts his strategy to improve his own payoff, he improves the payoff of Extortion even more. Another subclass of ZD called Generous also enforces a positive relation between two players' payoffs, but now the opponent's payoff is higher than the payoff of Generous \cite{Stewart2013PNAS}. In this case, as the opponent adapts to improve his own payoff, both players move toward mutual cooperation \cite{Stewart2013PNAS, Hilbe2018NatHumBehav}. Thus, facing ZD strategies with the positive slope of the line, when the opponent adjusts his own strategy to increase his payoff, he increases the ZD player's payoff even more. Moreover, when he achieves his own maximum payoff, the ZD player's payoff is also maximized. These strategies are called positively correlated ZD (pcZD) strategies\cite{ChenZinger2014JTheorBiol}.
These characteristics have attracted much attention from researchers in evolutionary games.
Thus, ZD strategies have been studied in the context of evolutionary games \cite{Akin2016ErgodicTheory, Adami2013NatComm, Hilbe2013PNAS, Hilbe2013PlosOne-zd, ChenZinger2014JTheorBiol, Szolnoki2014PhysRevE-zd, Szolnoki2014SciRep-zd, WuRong2014PhysRevE-zd, Hilbe2015JTheorBiol, LiuLi2015PhysicaA-zd, Xu2017PhysRevE-zd, Wang2019Chaos, Stewart2013PNAS, Mao2018EPL, Xu2019Neurocomputing}.
Besides evolutionary games, ZD strategies have been studied from various directions.
Examples are games with a discount factor \cite{Hilbe2015GamesEconBehav, Mcavoy2016PNAS, Mcavoy2017TheorPopulBiol, IchinoseMasuda2018JTheorBiol, GovaertCao2019arXiv}, games with observation errors \cite{Hao2015PhysRevE,MamiyaIchinose2019JTheorBiol}, multiplayer games \cite{Hilbe2014PNAS-zd,Hilbe2015JTheorBiol,Pan2015SciRep-zd,Milinski2016NatComm,Stewart2016PNAS, UedaTanaka2020PlosOne}, continuous action spaces \cite{Mcavoy2016PNAS,Milinski2016NatComm,Stewart2016PNAS,Mcavoy2017TheorPopulBiol}, alternating games \cite{Mcavoy2017TheorPopulBiol}, asymmetric games \cite{EngelFeigel2019ApplMathComput}, animal contests \cite{EngelFeigel2018PhysRevE}, human reactions to computerized ZD strategies \cite{Hilbe2014NatComm,Wang2016NatComm-zd}, and human-human experiments \cite{Hilbe2016PlosOne,Milinski2016NatComm, Becks2019NatComm}.

In those studies, Mamiya and Ichinose showed the existence of ZD strategies in games with a discount factor and observation errors \cite{MamiyaIchinose2020PhysRevE}.
These two factors are an important generalization because they are better able to capture real life interactions which are often noisy.
In their analyses, however, the conditions for the existence of zero-determinant strategies are not mathematically analyzed although some numerical examples are shown.
Here, we analytically derive those conditions and study how these two factors affect the existence of ZD strategies.

\section{Model}
%\subsection{RPD with private monitoring}
We consider the symmetric two-player two-action RPD game with imperfect private monitoring. % based on the literature \cite{Sekiguchi1997JEconTheor, Hao2015PhysRevE}. 
Each player $i\in\{X,Y\}$ chooses an action $a_i\in \{\rm{C},\rm{D}\}$ in each round, where C and D imply cooperation and defection, respectively. After the two players conduct the action, player $i$ observes  private signal $\omega_i\in\{g,b\}$ about the opponent's action, where $g$ and $b$ imply \textit{good} and \textit{bad}, respectively. 
Let $g$ ($b$) be the correct signal against the action C (D).  $\sigma(\bm \omega|\bm a)$ is the probability that a signal profile $\bm \omega=(\omega_X,\omega_Y)$ is realized  when the action profile is $\bm a=(a_X,a_Y)$ \cite{Sekiguchi1997JEconTheor}. Let $\epsilon$  be the probability that an error occurs to one particular player but not to the other player while $\xi$ be the probability that an error occurs to both players. Then, the probability that an error occurs to neither player is $\tau=1-2\epsilon-\xi$. We assume that the probability of observing a correct signal is higher than that of an incorrect one; $1/2 <\tau <1$.
For example, when both players take cooperation, $\sigma((g,g)|({\rm C},{\rm C}))=1-2\epsilon-\xi$, $\sigma((g,b)|({\rm C},{\rm C}))= \sigma((b,g)|(\rm{C},\rm{C}))=\epsilon$, and $\sigma((b,b)|({\rm C}, {\rm C}))=\xi$ are realized. 

In each round, player $i$'s realized payoff $u_i(a_i,\omega_i)$ is determined by his own action $a_i$ and signal $\omega_i$, such that $u_i({\rm C},g)=R$, $u_i({\rm C},b)=S$, $u_i({\rm D},g)=T$, and $u_i({\rm D},b)=P$.
%Note that the payoffs depend on the signals in private monitoring. 
Hence, player $i$'s expected payoff against the action profile $\bm a$ is given by
\begin{equation}
 f_i(\bm a)=\sum_{\bm \omega} u_i(a_i,\omega_i)\sigma(\bm \omega| \bm a).
 \label{cond_err}
\end{equation}
The expected payoff is determined by only action profile $\bm a$ regardless of signal profile $\bm \omega$. Thus, the expected payoff matrix against actions is given by
\begin{equation}
\bordermatrix{
 &  {\rm C} &  {\rm D} \cr
{\rm C} & R_E & S_E \cr
{\rm D} & T_E & P_E \cr}.% \;
\label{eq:payoff}
\end{equation}
According to Eq.~\eqref{cond_err}, $R_E$, $S_E$, $T_E$, and $P_E$ are derived as $R_E=R (1-\epsilon-\xi)+S (\epsilon+\xi)$, $S_E=S(1-\epsilon-\xi)+R(\epsilon+\xi)$, $T_E=T(1-\epsilon-\xi)+P(\epsilon+\xi)$, $P_E=P(1-\epsilon-\xi)+T(\epsilon+\xi)$, respectively. We assume that
\begin{equation}
T_E>R_E>P_E>S_E,
\label{eq:T>R>P>S}
\end{equation}
and
\begin{equation}
2R_E>T_E+S_E,
\label{eq:2R>T+S}
\end{equation}
which dictate the RPD conditions with imperfect private monitoring. In this paper, $T$, $R$, $P$, and $S$ are chosen such that $T_E$, $R_E$, $P_E$, and $S_E$ are fixed.

We introduce a discount factor to the repeated game. The game is to be played repeatedly over an infinite time horizon but the payoff will be discounted over rounds. Player $i$'s  discounted payoff of action profiles $\bm a(t)$, $t\in\mathbb{N}$ is $\delta^t f_i(\bm a(t))$ where $\delta$ is a discount factor and $t$ is a round.
$\delta$ can also be interpreted as the probability that the next round takes place \cite{Wang2015PhysLifeRev}.
%This game can be interpreted as repeated games with a finite but undetermined time horizon. 
Finally, the average discounted payoff of player $i$ is
\begin{equation}
s_i=(1-\delta)\sum_{t=0}^{\infty}\delta^tf_i(\bm a(t)).
\label{eq:s_i}
\end{equation}

Consider player $i$ that adopts memory-one strategies, with which they can use only the outcomes of the last round to decide the action to be submitted in the current round. A memory-one strategy is specified by a 5-tuple; $X$'s strategy is given by a combination of 
\begin{equation}
\label{eq:def bm p}
\bm p=(p_1,p_2,p_3,p_4;p_0),
\end{equation}
where $0\le p_j\le 1, j\in \{0,1,2,3,4\}$. The subscripts 1, 2, 3, and 4 of $p$ mean previous outcome C$g$, C$b$, D$g$ and D$b$, respectively. In Eq.~\eqref{eq:def bm p}, $p_{\rm 1}$ is the conditional probability that $X$ cooperates when $X$ cooperated and observed signal $g$ in the last round, $p_{\rm 2}$ is the conditional probability that $X$ cooperates when $X$ cooperated and observed signal $b$ in the last round, $p_{\rm 3}$ is the conditional probability that $X$ cooperates when $X$ defected and observed signal $g$ in the last round, and $p_{\rm 4}$ is the conditional probability that $X$ cooperates when $X$ defected and observed signal $b$ in the last round. Finally, $p_0$ is the probability that $X$ cooperates in the first round. Similarly, $Y$'s strategy is specified by a combination of 
\begin{equation}
\bm q=(q_1,q_2,q_3,q_4;q_0),
\end{equation}
where $0\le q_j\le 1, j\in \{0,1,2,3,4\}$. 

Define $\bm v(t)=(v_1(t),v_2(t),v_3(t),v_4(t))$ as the stochastic state of two players in round $t$ where the subscripts 1, 2, 3, and 4 of $v$ imply the stochastic states (C,C), (C,D), (D,C), and (D,D), respectively.
$v_1(t)$ is the probability that both players cooperate in round $t$, $v_2(t)$ is the probability that $X$ cooperates and $Y$ defects in round $t$, and so forth. Then, the expected payoff to player $X$ in round $t$ is given by $\bm v(t) \bm S_X$, where $\bm S_X^T=(R_E,S_E,T_E,P_E)$.
The expected per-round payoff to player $X$ in the repeated game is given by
\begin{equation}
s_X=(1-\delta)\sum_{t=0}^{\infty}\delta^t \bm v(t) \bm S_X,
\label{eq:s_x}
\end{equation}
where $0<\delta<1$. 

%Eq.~\eqref{eq:s_x} can be represented by a determinant form. 
The initial stochastic state is given by
\begin{equation}
	\bm v(0)=(p_0q_0,p_0(1-q_0),(1-p_0)q_0,(1-p_0)(1-q_0)).
\end{equation}
The state transition matrix $M$ of these repeated games with observation errors is given by
\begin{equation}
\label{eq:M}
M=
\scalebox{0.85}{$\displaystyle
 \left(
 \begin{array}{ll}
    \left(
    \begin{array}{ll}
      \tau p_1 q_1  \\
      +\epsilon p_1 q_2 \\
      +\epsilon p_2 q_1 \\
      +\xi p_2 q_2 
    \end{array}
    \right)
    \left(
    \begin{array}{ll}
      \tau p_1 (1-q_1)  \\
      +\epsilon p_1 (1-q_2) \\
      +\epsilon p_2 (1-q_1) \\
      +\xi p_2 (1-q_2) 
    \end{array}
    \right)
    \left(
    \begin{array}{ll}
      \tau (1-p_1) q_1  \\
      +\epsilon (1-p_1) q_2 \\
      +\epsilon (1-p_2) q_1 \\
      +\xi (1-p_2) q_2 
    \end{array}
    \right)
    \left(
    \begin{array}{ll}
      \tau (1-p_1) (1-q_1)  \\
      +\epsilon (1-p_1) (1-q_2) \\
      +\epsilon (1-p_2) (1-q_1) \\
      +\xi (1-p_2) (1-q_2) 
    \end{array}
    \right)\\
  \left(
    \begin{array}{ll}
      \epsilon p_1 q_3  \\
      +\xi p_1 q_4 \\
      +\tau p_2 q_3 \\
      +\epsilon p_2 q_4 
    \end{array}
    \right)
    \left(
    \begin{array}{ll}
      \epsilon p_1 (1-q_3)  \\
      +\xi p_1 (1-q_4) \\
      +\tau p_2 (1-q_3) \\
      +\epsilon p_2 (1-q_4) 
    \end{array}
    \right)
    \left(
    \begin{array}{ll}
      \epsilon (1-p_1) q_3  \\
      +\xi (1-p_1) q_4 \\
      +\tau (1-p_2) q_3 \\
      +\epsilon (1-p_2) q_4 
    \end{array}
    \right)
    \left(
    \begin{array}{ll}
      \epsilon (1-p_1) (1-q_3)  \\
      +\xi (1-p_1) (1-q_4) \\
      +\tau (1-p_2) (1-q_3) \\
      +\epsilon (1-p_2) (1-q_4) 
    \end{array}
    \right)\\
    \left(
    \begin{array}{ll}
      \epsilon p_3 q_1  \\
      +\tau p_3 q_2 \\
      +\xi p_4 q_1 \\
      +\epsilon p_4 q_2 
    \end{array}
    \right)
    \left(
    \begin{array}{ll}
      \epsilon p_3 (1-q_1)  \\
      +\tau p_3 (1-q_2) \\
      +\xi p_4 (1-q_1) \\
      +\epsilon p_4 (1-q_2) 
    \end{array}
    \right)
    \left(
    \begin{array}{ll}
      \epsilon (1-p_3) q_1  \\
      +\tau (1-p_3) q_2 \\
      +\xi (1-p_4) q_1 \\
      +\epsilon (1-p_4) q_2 
    \end{array}
    \right)
    \left(
    \begin{array}{ll}
      \epsilon (1-p_3) (1-q_1)  \\
      +\tau (1-p_3) (1-q_2) \\
      +\xi (1-p_4) (1-q_1) \\
      +\epsilon (1-p_4) (1-q_2) 
    \end{array}
    \right)\\
    \left(
    \begin{array}{ll}
      \xi p_3 q_3  \\
      +\epsilon p_3 q_4 \\
      +\epsilon p_4 q_3 \\
      +\tau p_4 q_4 
    \end{array}
    \right)
    \left(
    \begin{array}{ll}
      \xi p_3 (1-q_3)  \\
      +\epsilon p_3 (1-q_4) \\
      +\epsilon p_4 (1-q_3) \\
      +\tau p_4 (1-q_4) 
    \end{array}
    \right)
    \left(
    \begin{array}{ll}
      \xi (1-p_3) q_3  \\
      +\epsilon (1-p_3) q_4 \\
      +\epsilon (1-p_4) q_3 \\
      +\tau (1-p_4) q_4 
    \end{array}
    \right)
    \left(
    \begin{array}{ll}
      \xi (1-p_3) (1-q_3)  \\
      +\epsilon (1-p_3) (1-q_4) \\
      +\epsilon (1-p_4) (1-q_3) \\
      +\tau (1-p_4) (1-q_4) 
    \end{array}
    \right)
 \end{array}
  \right)
$}.
\end{equation}
%where $\tau =1-2\epsilon-\xi$. 
Then, we obtain 
\begin{equation}
s_X=(1-\delta)\bm v(0)\sum_{t=0}^{\infty} (\delta M)^t \bm S_X=\bm v^T \bm S_X,
\label{eq:s_x_IM}
\end{equation}
where $\bm v^T \equiv (1-\delta){\bm v(0)}(I-\delta M)^{-1}$, which is the mean distribution of $\bm v(t)$ and $I$ is the $4\times 4$ identity matrix. Additionally, we define
\begin{equation}\label{def:M0}
  M_0 = \left(
    \begin{array}{cccc}
      p_0 q_0 & p_0(1-q_0) & (1-p_0)q_0 &(1-p_0)(1-q_0)\\
      p_0 q_0 & p_0(1-q_0) & (1-p_0)q_0 &(1-p_0)(1-q_0)\\
      p_0 q_0 & p_0(1-q_0) & (1-p_0)q_0 &(1-p_0)(1-q_0)\\
      p_0 q_0 & p_0(1-q_0) & (1-p_0)q_0 &(1-p_0)(1-q_0)
    \end{array}
  \right).
\end{equation}
Because $v_1+v_2+v_3+v_4=1$, $\bm v(0)=\bm v^TM_0$ holds. By substituting $\bm v(0)=\bm v^TM_0$ in $\bm v^T=(1-\delta){\bm v(0)}(I-\delta M)^{-1}$ and multiplying  both sides of the equation by $(I-\delta M)$ from the right, we obtain $\bm v^T(I-\delta M)=(1-\delta)\bm v^T M_0$. Thus, we obtain $\bm v^TM^\prime=\bm 0$, where $M^\prime\equiv \delta M+(1-\delta)M_0-I$. We immediately obtain an expression for the dot product of an arbitrary vector $\bm f^T=(f_1,f_2,f_3,f_4)$ with the fourth column vector $\bm u$ of matrix $M^\prime$ as a consequence of Press and Dyson's formalism, which can be represented by the form of the determinant
\begin{equation}\label{eq:D_err}
\begin{split}
&\bm u\cdot \bm f=\\
&
\scalebox{0.85}{$
  \left|
    \begin{array}{cccc}
      \delta(\tau p_1 q_1 +\epsilon p_1 q_2 +\epsilon p_2 q_1+\xi p_2 q_2) -1 +p_0 q_0(1-\delta)& \delta(\mu p_{1}+\eta p_{2})-1 +p_0(1-\delta)& \delta(\mu q_{1}+\eta q_{2})-1 +q_0 (1-\delta)& f_1\\
      \delta(\epsilon p_1 q_3+\xi p_1 q_4+\tau p_2 q_3+\epsilon p_2 q_4)       +p_0 q_0(1-\delta)& \delta(\eta p_{1}+\mu p_{2})-1 +p_0(1-\delta)& \delta(\mu q_{3}+\eta q_{4})   +q_0 (1-\delta)& f_2\\
      \delta(\epsilon p_3 q_1+\tau p_3 q_2+\xi p_4 q_1+\epsilon p_4 q_2)       +p_0 q_0(1-\delta)& \delta(\mu p_{3}+\eta p_{4})   +p_0(1-\delta)&  \delta(\eta q_{1}+\mu q_{2})-1+q_0 (1-\delta)& f_3\\      
      \delta(\xi p_3 q_3+\epsilon p_3 q_4+\epsilon p_4 q_3+\tau p_4 q_4)       +p_0 q_0(1-\delta)& \delta(\eta p_{3}+\mu p_{4})   +p_0(1-\delta)& \delta(\eta q_{3}+\mu q_{4})   +q_0 (1-\delta)& f_4
    \end{array}
  \right|
$}\\
& \equiv D(\bm p,\bm q,\bm f),
\end{split}
\end{equation}
where $\mu=1-\epsilon-\xi$ and $\eta=\epsilon+\xi$. Furthermore, Eq.~\eqref{eq:D_err} should be normalized to have its components sum to $1$ by $\bm u \cdot \bm 1$, where ${\bm 1}=(1,1,1,1)$. Then, we obtain the dot product of an arbitrary vector $\bm f$ with  mean distribution $\bm v$. %Replacing the last column of $D(\bm p,\bm q, \bm f)$ with player $X$'s and $Y$'s expected payoff vector, respectively, 
Therefore, we obtain their per-round expected payoffs:
\begin{equation}
s_X=\bm v \cdot \bm S_X=\frac{\bm u \cdot \bm S_X}{\bm u \cdot \bm 1}
   =\frac{D(\bm p,\bm q,\bm S_X)}{D(\bm p,\bm q,\bm 1)},
\label{eq:s_x_Det}
\end{equation}
\begin{equation}
s_Y=\bm v \cdot \bm S_Y=\frac{\bm u \cdot \bm S_Y}{\bm u \cdot \bm 1}
   =\frac{D(\bm p,\bm q,\bm S_Y)}{D(\bm p,\bm q,\bm 1)}.
\label{eq:s_y_Det}
\end{equation}
When we set $\delta=1$, Eq.~\eqref{eq:D_err} corresponds to Eq.~(2) of \cite{Hao2015PhysRevE}. 
By using Eq.~\eqref{eq:D_err}, we can calculate players' per-round expected payoffs when $0< \delta \le 1$ by the form of the determinants.
$\delta=1$ is the case where future payoffs are not discounted.

\section{Result}
We analytically study conditions for the existence of Equalizer and pcZD strategies as a function of observation errors and a discount factor in Repeated Prisoner's Dilemma games with imperfect private monitoring.
%  that lead the opponent to unconditionally cooperate

A player $X$ can choose a strategy $\bm p$ so that he unilaterally enforces $\alpha s_X+\beta s_Y+\gamma=0$ against  Eqs.~\eqref{eq:s_x_Det} and \eqref{eq:s_y_Det} for any opponent's strategy $\bm q$. As shown in \cite{MamiyaIchinose2020PhysRevE}, such a strategy of player $X$ is given by
\begin{equation}\label{ZD_ErrorDiscoutFactor}
\begin{split}
      \delta(\mu  p_1+\eta p_2)-1+p_0(1-\delta)&=\alpha R_E +\beta R_E+\gamma,\\
      \delta(\eta p_1+\mu  p_2)-1+p_0(1-\delta)&=\alpha S_E +\beta T_E+\gamma,\\
      \delta(\mu  p_3+\eta p_4)+p_0(1-\delta)  &=\alpha T_E +\beta S_E+\gamma,\\
      \delta(\eta p_3+\mu  p_4)+p_0(1-\delta)  &=\alpha P_E +\beta P_E+\gamma.
\end{split}
\end{equation}
Strategies which satisfy Eq.~\eqref{ZD_ErrorDiscoutFactor} are ZD strategies. $\alpha$, $\beta$, and $\gamma$ are parameters determined by player $X$. Equalizer, Extortion, Generous, and pcZD strategies are derived by giving appropriate values to $\alpha$, $\beta$, and $\gamma$. In this paper, $\eta$ satisfies $\eta=\epsilon+\xi<1/2$ because of assuming $1/2<\tau<1$.

\subsection{Equalizer}
\subsubsection{Expression}
We substitute $\alpha =0$ into Eq.~\eqref{ZD_ErrorDiscoutFactor} to obtain Equalizer:
\begin{equation}\label{equalizer_error}
\begin{split}
      \delta(\mu  p_1+\eta p_2)-1+p_0(1-\delta)&=\beta R_E+\gamma,\\
      \delta(\eta p_1+\mu  p_2)-1+p_0(1-\delta)&=\beta T_E+\gamma,\\
      \delta(\mu  p_3+\eta p_4)+p_0(1-\delta)  &=\beta S_E+\gamma,\\
      \delta(\eta p_3+\mu  p_4)+p_0(1-\delta)  &=\beta P_E+\gamma.
\end{split}
\end{equation}
Equalizer can fix the opponent's payoff no matter what the opponent takes, which means that $s_Y=-\gamma/\beta$. From the four equations in Eq.~\eqref{equalizer_error} we have
\begin{equation}\label{eq:equalizer_beta-gamma}
\begin{split}
\beta &=
 -\frac{(1-\delta p_1+\delta p_4)(\mu-\eta)}
 {\mu(R_E-P_E)-\eta(T_E-S_E)},\\
\gamma&=
 \frac{(1-\delta p_1-p_0+\delta p_0)(\mu P_E-\eta S_E)+(p_0-\delta p_0+\delta p_4)(\mu R_E-\eta T_E)}
 {\mu(R_E-P_E)-\eta(T_E-S_E)}.
\end{split}
\end{equation}
Also, $p_2$ and $p_3$ can be rewritten  as
\begin{equation}\label{eq:equalizer_p2-p3}
\begin{split}
p_2&= \frac{p_1\left(\mu(T_E-P_E)-\eta(R_E-S_E)\right)-(\frac{1}{\delta}+p_4)(T_E-R_E)}{\mu(R_E-P_E)-\eta(T_E-S_E)},\\
p_3&= \frac{(\frac{1}{\delta}-p_1)(P_E-S_E)+p_4\left(\mu(R_E-S_E)-\eta(T_E-P_E)\right)}{\mu(R_E-P_E)-\eta(T_E-S_E)}.
\end{split}
\end{equation}
We substitute Eq.~\eqref{eq:equalizer_beta-gamma} into $s_Y=-\gamma/\beta$ to obtain 
\begin{equation}\label{eq:equal_sy}
	s_Y=\frac{(1-\delta p_1-p_0+\delta p_0)(\mu P_E-\eta S_E)+(p_0-\delta p_0+\delta p_4)(\mu R_E-\eta T_E)}{(1-\delta p_1+\delta p_4)(\mu-\eta)}.
\end{equation}
Eq.~\eqref{eq:equal_sy} is a weighed average of $(\mu P_E-\eta S_E)/(\mu-\eta)$ and $(\mu R_E-\eta T_E)/(\mu-\eta)$ with non-negative weights. Therefore, when $\eta$ satisfies $\mu(R_E-P_E)-\eta(T_E-S_E)>0$, that is
\begin{equation}\label{eta_c}
 \eta < \eta_c\equiv \frac{R_E-P_E}{T_E+R_E-P_E-S_E},
\end{equation}
Equalizer can impose any payoff value $s_Y$ such that
\begin{equation}
\frac{\mu P_E-\eta S_E}{\mu-\eta}\le s_Y\le \frac{\mu R_E-\eta T_E}{\mu-\eta}.
\end{equation}
If player $X$ sets $p_4=p_0=0$, he can impose $s_Y=(\mu P_E -\eta S_E)/(\mu-\eta)$. If player $X$ sets $p_1=p_0=1$, he can impose $s_Y=(\mu R_E -\eta T_E)/(\mu-\eta)$. In the previous study \cite{MamiyaIchinose2020PhysRevE}, the condition of error rates for the existence of Equalizer was not analytically obtained. In this study, we analytically derived the condition, that is Eq.~\eqref{eta_c}. Any Equalizers no longer exist when $\eta$ satisfies $\eta_c\le \eta<1/2$ (Appendix~\ref{noLongerExist_Equalizer}).

\subsubsection{Existence condition}

In this section, we identify the condition for $\delta$ under which Equalizer can exist. When $\eta$ satisfies $\eta < \eta_c$, $\eta$ also satisfies $\mu(T_E-P_E)-\eta(R_E-S_E)>0$ and $\mu(R_E-S_E)-\eta(T_E-P_E)>0$ because of the RPD condition $T_E>R_E>P_E>S_E$. As you can see in Appendix~\ref{noLongerExist_Equalizer}, any Equalizers no longer exist when $\eta$ satisfies $\eta_c\le \eta<1/2$.  Eq.~\eqref{eq:equalizer_p2-p3} indicates that an Equalizer strategy exists if and only if
\begin{equation}\label{eq:inequality_p3}
0\le 
p_1(\mu(T_E-P_E)-\eta(R_E-S_E))-(\frac{1}{\delta}+p_4)(T_E-R_E)
\le \mu(R_E-P_E)-\eta(T_E-S_E),
\end{equation}
\begin{equation}\label{eq:inequality_p4}
0\le 
(\frac{1}{\delta}-p_1)(P_E-S_E)+p_4(\mu(R_E-S_E)-\eta(T_E-P_E))
\le \mu(R_E-P_E)-\eta(T_E-S_E),
\end{equation}
for some $0\le p_1\le 1$ and $0\le p_4\le 1$. Independent of $\delta$, $p_1$ and $p_4$ satisfies the second inequality of Eq.~\eqref{eq:inequality_p3} and the first inequality of Eq.~\eqref{eq:inequality_p4} because they are satisfied in the most stringent case, i.e, $p_1=1$ and $p_4=0$. The first inequality of Eq.~\eqref{eq:inequality_p3} and the second inequality of Eq.~\eqref{eq:inequality_p4} lead
\begin{equation}
p_4\le \frac{\mu(T_E-P_E)-\eta(R_E-S_E)}{T_E-R_E} p_1 -\frac{1}{\delta},
\end{equation}
and
\begin{equation}
p_4 
\le \frac{\mu(R_E-P_E)-\eta(T_E-S_E)}{\mu(R_E-S_E)-\eta(T_E-P_E)}
+(p_1-\frac{1}{\delta})\frac{P_E-S_E}{\mu(R_E-S_E)-\eta(T_E-P_E)},
\end{equation}
respectively. From these inequalities, the feasible set $(p_1, p_4)$ becomes smaller as $\delta$ decreases. When $\delta$ decreases, the last Equalizer left is the strategy with $p_1=1$ and $p_4=0$. Therefore, the condition for $\delta$ under which Equalizer strategies exist is given by
\begin{equation}\label{delta_c}
 \delta \ge \delta_c\equiv \max\left(\frac{T_E-R_E}{\mu(R_E-P_E)-\eta(T_E-S_E)+T_E-R_E},\frac{P_E-S_E}{\mu(R_E-P_E)-\eta(T_E-S_E)+P_E-S_E}\right).
\end{equation}
In the case that $R_E+P_E<T_E+S_E$ is satisfied, the first element in Eq.~\eqref{delta_c} is larger than the second one (Eq.~\eqref{eq:TR_PS} of Appendix~\ref{ap:delta}). In the case that $R_E+P_E>T_E+S_E$ is satisfied, the second element in Eq.~\eqref{delta_c} is larger than the first one (Eq.~\eqref{eq:PS_TR} of Appendix~\ref{ap:delta}). Finally, in the case that $R_E+P_E=T_E+S_E$ is satisfied, both elements take the same value.

When there are no observation errors but a discount factor ($\eta=0$ and $\delta<1$), $\delta_c$ with $\eta=0$ corresponds to Eq.~(32) in Ref.~\cite{IchinoseMasuda2018JTheorBiol}.

\subsubsection{Numerical examples}
\begin{figure}[h]
  \centering
  \includegraphics[width=0.8\columnwidth]{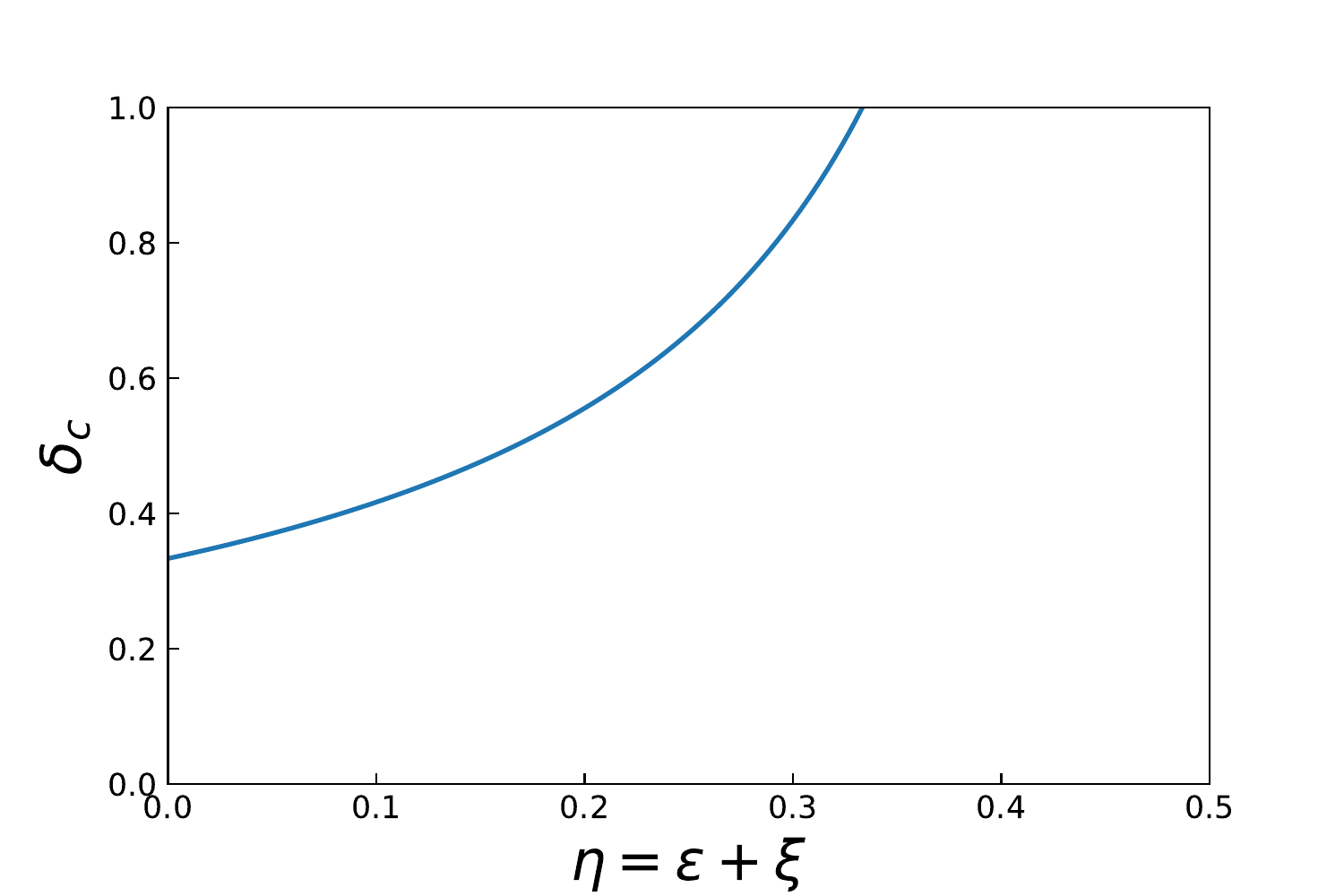}
  \caption{Minimum discount factor $\delta_c$ against error rates $\eta$.}
  \label{figure1_1}
\end{figure}

Here we numerically show the minimum discount factor $\delta_c$ against error rates (Fig.~\ref{figure1_1}).
We used $(T_E,R_E,P_E,S_E)=(1.5,1,0,-0.5)$ in the figure.
As the figure shows, $\delta_c=1/3$ when there are no errors ($\eta=0$).
Thus, Equalizer can exist within the range of $1/3 \leq \delta \leq 1$.
However, as error rates increase, $\delta_c$ becomes larger.
Large error rates make the range of $\delta$ narrow where Equalizer can exist.
As shown in the figure, $\delta_c=0.4167$ when $\eta=0.1$, $\delta_c=0.5556$ when $\eta=0.2$, and $\delta_c=0.8333$ when $\eta=0.3$.
When $\eta=1/3$, Equalizer can only exist at $\delta=1$.

\begin{figure}[h]
  \centering
  \includegraphics[width=0.8\columnwidth]{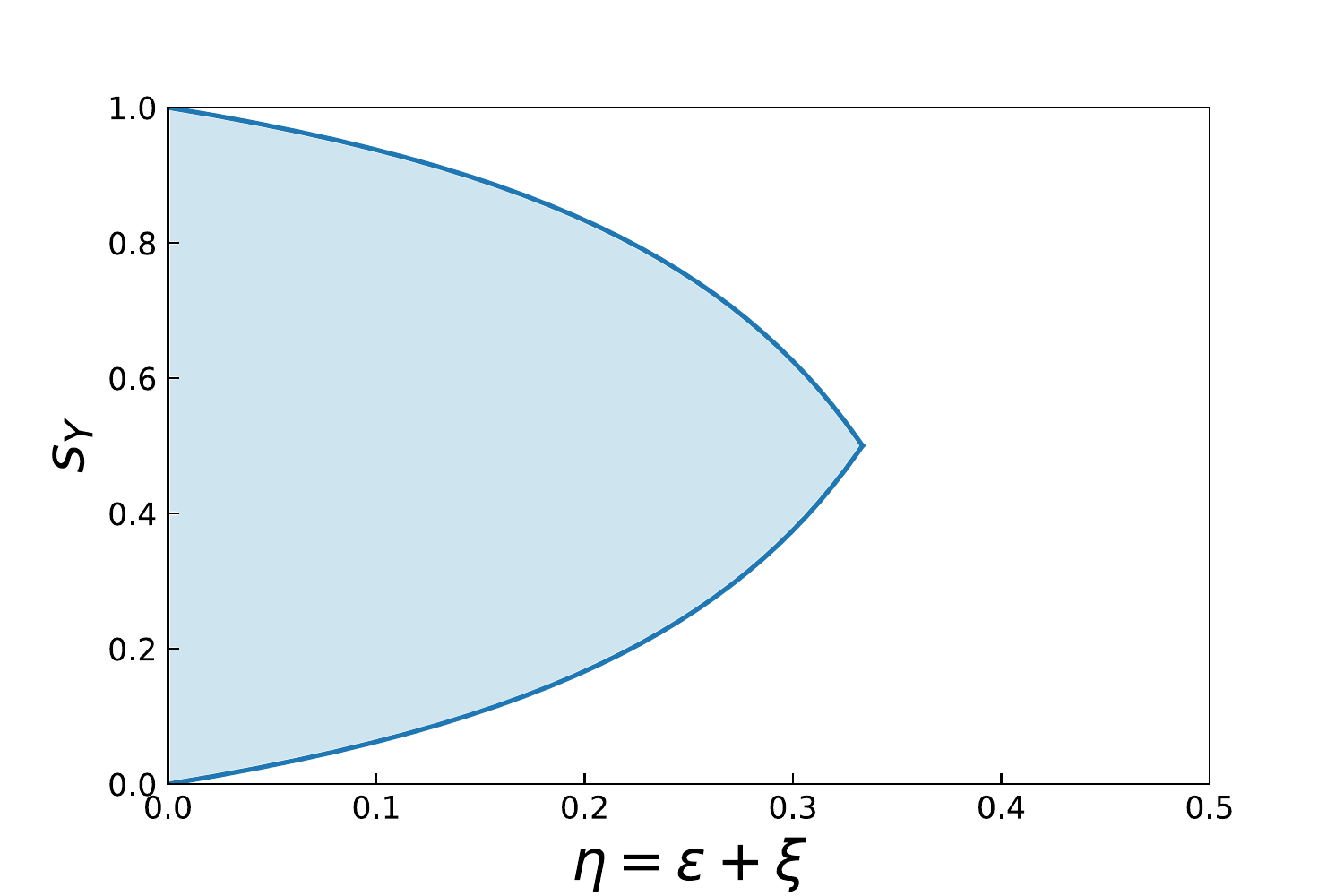}
  \caption{Ranges of expected opponent's payoffs $s_Y$ which Equalizer can enforce.}
  \label{figure1_2}
\end{figure}

Figure \ref{figure1_2} shows the range of the expected payoff of the opponent which Equalizer can enforce as a function of error rates.
When there are no errors, Equalizer can fix the opponent's expected payoff to any values between $P_E = 0$ and $R_E = 1$.
However, when there are errors, the controllability of Equalizer becomes weak.
As error rates increase, the range of the opponent's payoff becomes narrower.
The ranges of the expected opponent's payoffs are $0.0625 \le s_Y \le 0.9375$ when $\eta=0.1$, $0.1667 \le s_Y \le 0.8333$ when $\eta=0.2$, and $0.3750 \le s_Y \le 0.6250$ when $\eta=0.3$.
When $\eta=1/3$, Equalizer can only fix the opponent's payoff at $s_Y=0.5$.
When $\eta>1/3$, Equalizer no longer exists.

\subsection{Positively correlated ZD strategies}
\subsubsection{Expression}
By substituting $\alpha=\phi$, $\beta=-\phi \chi$, and $\gamma=\phi (\chi-1)\kappa$ into Eq.~\eqref{ZD_ErrorDiscoutFactor}, ZD strategies for player $X$ can be rewritten  as
\begin{equation}\label{eq:ZDEq}
\begin{split}
      \delta(\mu  p_1+\eta p_2) + (1-\delta)p_0&=1-\phi(\chi-1)(R_E-\kappa),\\
      \delta(\eta p_1+\mu  p_2) + (1-\delta)p_0&=1-\phi
      \left(T_E-S_E+(\chi-1)(T_E-\kappa) \right),\\
      \delta(\mu  p_3+\eta p_4) + (1-\delta)p_0&=
      \phi\left(T_E-S_E+(\chi-1)(\kappa-S_E) \right),\\
      \delta(\eta p_3+\mu  p_4) + (1-\delta)p_0&=\phi(\chi-1)(\kappa-P_E),
\end{split}
\end{equation}
where $\phi$, $\chi$, and $\kappa$ are the coefficients. Equations~\eqref{eq:s_x_Det} and \eqref{eq:s_y_Det} allow player $X$ using ZD strategies to enforce 
\begin{equation}\label{eq:linear_eq_kappa_chi}
s_X - \kappa = \chi (s_Y - \kappa)
\end{equation}
for any strategy $\bm q$ of the opponent player $Y$. We call $\chi$ a correlation factor. The parameter $\kappa$ is called the baseline payoff. Player $X$ can make the payoffs between $X$ and $Y$ positively correlated with $\chi \ge 1$. ZD strategies which enforce Eq.~\eqref{eq:linear_eq_kappa_chi} with some fixed $\chi \ge 1$ are called positively correlated ZD (pcZD) strategies in Chen and Zinger \cite{ChenZinger2014JTheorBiol}.

\subsubsection{Existence condition}
\label{Mathematical Analysis}
In this section, we analytically derive the condition for pcZD strategies to exist according to the degree of error rates and a discount factor.
When $\delta<1$ and $\chi=1$ hold, no solution $p_j, j\in \{0,1,2,3,4\}$ which satisfies Eq.~\eqref{eq:ZDEq} exists. Thus, ZD strategies with $\chi=1$ do not exist when $\delta<1$. Therefore, in the following, we assume $\chi \neq1$. By solving Eq.~\eqref{eq:ZDEq} for $p_j$, $j\in \{1, 2, 3, 4\}$, 
\begin{equation}\label{eq:ZDEq2}
\begin{split}
\delta p_1 +(1-\delta)p_0&=1-\phi \hat{\chi}
\left \{R_E-\kappa-\frac{\eta}{\mu-\eta}\left(\frac{T_E-S_E}{\hat{\chi}}+T_E-R_E\right)\right\},\\
\delta p_2 +(1-\delta)p_0&=1-\phi \hat{\chi}
\left\{R_E-\kappa+\frac{\mu}{\mu-\eta}\left(\frac{T_E-S_E}{\hat{\chi}}+T_E-R_E\right)\right\},\\
\delta p_3 +(1-\delta)p_0&= \phi \hat{\chi}
\left\{\kappa-P_E+\frac{\mu}{\mu-\eta}\left(\frac{T_E-S_E}{\hat{\chi}}+P_E-S_E\right)\right\},\\
\delta p_4 +(1-\delta)p_0&=\phi \hat{\chi}
\left\{\kappa-P_E -\frac{\eta}{\mu-\eta}\left(\frac{T_E-S_E}{\hat{\chi}}+P_E-S_E\right)\right\},
\end{split}
\end{equation}
are obtained, where $\hat{\chi}=\chi-1\neq0$. By applying $0\le p_j\le 1$, $j\in \{1, 2, 3, 4\}$ to Eq.~\eqref{eq:ZDEq2}
\begin{equation}\label{eq:Inequality1}
(1-\delta)(1-p_0)\le
\phi \hat{\chi}\left\{R_E-\kappa-\frac{\eta}{\mu-\eta}\left(\frac{T_E-S_E}{\hat{\chi}}+T_E-R_E\right) \right\}
\le(1-\delta)(1-p_0)+\delta,
\end{equation}
\begin{equation}\label{eq:Inequality2}
(1-\delta)(1-p_0)
\le \phi \hat{\chi}\left\{R_E-\kappa+\frac{\mu}{\mu-\eta}\left(\frac{T_E-S_E}{\hat{\chi}}+T_E-R_E\right)\right\}
\le(1-\delta)(1-p_0)+\delta,
\end{equation}
\begin{equation}\label{eq:Inequality3}
(1-\delta)p_0\le 
\phi \hat{\chi}\left\{\kappa-P_E+\frac{\mu}{\mu-\eta}\left(\frac{T_E-S_E}{\hat{\chi}}+P_E-S_E\right)\right\} 
\le(1-\delta)p_0+\delta,
\end{equation}
\begin{equation}\label{eq:Inequality4}
(1-\delta)p_0\le
\phi \hat{\chi}\left\{\kappa-P_E-\frac{\eta}{\mu-\eta}\left(\frac{T_E-S_E}{\hat{\chi}}+P_E-S_E\right)\right\}
\le(1-\delta)p_0+\delta,
\end{equation}
are obtained.
The sum of the first inequality in Eq.~\eqref{eq:Inequality1} multiplied by $\mu$ and the first inequality in Eq.~\eqref{eq:Inequality2} multiplied by $\eta$ becomes
\begin{equation}\label{phi_RE_kappa}
(1-\delta)(1-p_0)\le \phi (\chi-1)(R_E-\kappa).
\end{equation} 
Similarly, the sum of the first inequality in Eq.~\eqref{eq:Inequality3} multiplied by $\eta$ and the first inequality in Eq.~\eqref{eq:Inequality4} multiplied by $\mu$ becomes
\begin{equation}\label{phi_kappa_PE}
(1-\delta)p_0\le \phi (\chi-1)(\kappa-P_E).
\end{equation}
Summing up the obtained two inequalities shows $1-\delta\le \phi (\chi-1)(R_E-P_E)$. In particular, $\phi (\chi-1)>0$, therefore $\chi>1$ and $\phi>0$. $\phi<0$ and $\chi<1$ is omitted because if we assume $\chi<1$, no ZD strategies with $0<\chi<1$ exist and ZD strategies with $\chi \le 0$ are not pcZD strategies (Appendix~\ref{ap:chilessthan0}). Additionally, Eq.~\eqref{phi_RE_kappa} and Eq.~\eqref{phi_kappa_PE} imply $P_E\le\kappa\le R_E$ when $\phi (\chi-1)>0$.

Because all of the middle terms from Eq.~\eqref{eq:Inequality1} to Eq.~\eqref{eq:Inequality4} are greater than or equal to zero, we obtain 
\begin{equation}\label{cond_kappa_errorNoDiscount1}
P_E+\frac{\eta}{\mu-\eta}\left(\frac{T_E-S_E}{\hat{\chi}}+P_E-S_E\right)
\le\kappa \le R_E-\frac{\eta}{\mu-\eta}\left(\frac{T_E-S_E}{\hat{\chi}}+T_E-R_E\right).
\end{equation}

In the following, we analyze Eqs.~\eqref{eq:Inequality1}--\eqref{eq:Inequality4} by dividing into three cases (1) $0<p_0<1$, (2) $p_0=0$, (3) $p_0=1$ as follows.

{\bf (1) Case of $0<p_0<1$:}\\
Equations \eqref{eq:Inequality1}--\eqref{eq:Inequality4} lead to
\begin{equation}
\label{phi_cond_original1}
\frac{\hat{\chi}\left\{R_E-\kappa-\frac{\eta}{\mu-\eta}\left(\frac{T_E-S_E}{\hat{\chi}}+T_E-R_E\right) \right\}}{(1-\delta)(1-p_0)+\delta}
\le\frac{1}{\phi}\le
\frac{\hat{\chi}\left\{R_E-\kappa-\frac{\eta}{\mu-\eta}\left(\frac{T_E-S_E}{\hat{\chi}}+T_E-R_E\right) \right\}}{(1-\delta)(1-p_0)},
\end{equation}
\begin{equation}
\label{phi_cond_original2}
\frac{\hat{\chi}\left\{R_E-\kappa+\frac{\mu}{\mu-\eta}\left(\frac{T_E-S_E}{\hat{\chi}}+T_E-R_E\right)\right\}}{(1-\delta)(1-p_0)+\delta}
\le\frac{1}{\phi}
\le \frac{\hat{\chi}\left\{R_E-\kappa+\frac{\mu}{\mu-\eta}\left(\frac{T_E-S_E}{\hat{\chi}}+T_E-R_E\right)\right\}}{(1-\delta)(1-p_0)},
\end{equation}
\begin{equation}
\label{phi_cond_original3}
\frac{\hat{\chi}\left\{\kappa-P_E+\frac{\mu}{\mu-\eta}\left(\frac{T_E-S_E}{\hat{\chi}}+P_E-S_E\right)\right\}}{(1-\delta)p_0+\delta} 
\le\frac{1}{\phi}
\le\frac{\hat{\chi}\left\{\kappa-P_E+\frac{\mu}{\mu-\eta}\left(\frac{T_E-S_E}{\hat{\chi}}+P_E-S_E\right)\right\}}{(1-\delta)p_0},
\end{equation}
\begin{equation}
\label{phi_cond_original4}
\frac{\hat{\chi}\left\{\kappa-P_E-\frac{\eta}{\mu-\eta}\left(\frac{T_E-S_E}{\hat{\chi}}+P_E-S_E\right)\right\}}{(1-\delta)p_0+\delta}
\le\frac{1}{\phi}\le
\frac{\hat{\chi}\left\{\kappa-P_E-\frac{\eta}{\mu-\eta}\left(\frac{T_E-S_E}{\hat{\chi}}+P_E-S_E\right)\right\}}{(1-\delta)p_0}.
\end{equation}
The condition under which a positive $\phi$ value that satisfies Eqs.~\eqref{phi_cond_original1}--\eqref{phi_cond_original4} exists is given by
\begin{equation}
\label{phi_cond1}
\frac{\hat{\chi}\left\{R_E-\kappa+\frac{\mu}{\mu-\eta}\left(\frac{T_E-S_E}{\hat{\chi}}+T_E-R_E\right)\right\}}
{(1-\delta)(1-p_0)+\delta}
\le
\frac{\hat{\chi}\left\{R_E-\kappa-\frac{\eta}{\mu-\eta}\left(\frac{T_E-S_E}{\hat{\chi}}+T_E-R_E\right)\right\}}{(1-\delta)(1-p_0)},
\end{equation}
\begin{equation}\label{phi_cond2}
\frac{\hat{\chi}\left\{R_E-\kappa+\frac{\mu}{\mu-\eta}\left(\frac{T_E-S_E}{\hat{\chi}}+T_E-R_E\right)\right\}}
{(1-\delta)(1-p_0)+\delta}
\le
\frac{\hat{\chi}\left\{\kappa-P_E-\frac{\eta}{\mu-\eta}\left(\frac{T_E-S_E}{\hat{\chi}}+P_E-S_E\right)\right\}}
{(1-\delta)p_0},
\end{equation}
\begin{equation}\label{phi_cond3}
\frac{\hat{\chi}\left\{\kappa-P_E+\frac{\mu}{\mu-\eta}\left(\frac{T_E-S_E}{\hat{\chi}}+P_E-S_E\right)\right\}}
{(1-\delta)p_0+\delta}
\le
\frac{\hat{\chi}\left\{R_E-\kappa-\frac{\eta}{\mu-\eta}\left(\frac{T_E-S_E}{\hat{\chi}}+T_E-R_E\right)\right\}}
{(1-\delta)(1-p_0)},
\end{equation}
\begin{equation}\label{phi_cond4}
\frac{\hat{\chi}\left\{\kappa-P_E+\frac{\mu}{\mu-\eta}\left(\frac{T_E-S_E}{\hat{\chi}}+P_E-S_E\right)\right\}}
{(1-\delta)p_0+\delta}
\le
\frac{\hat{\chi}\left\{\kappa-P_E-\frac{\eta}{\mu-\eta}\left(\frac{T_E-S_E}{\hat{\chi}}+P_E-S_E\right)\right\}}
{(1-\delta)p_0}.
\end{equation}
Solving the above inequalities for $\kappa$, we obtain
\begin{equation}
\label{cond_kappa_ErrorDiscount1}
\kappa \le R_E-\frac{(\frac{1}{\delta}-1)(1-p_0)+\eta}{\mu-\eta}\left(\frac{T_E-S_E}{\hat{\chi}}+T_E-R_E\right),
\end{equation}
\begin{equation}\label{cond_kappa_ErrorDiscount2}
\kappa\ge P_E+\frac{\eta}{\mu-\eta}\left(\frac{T_E-S_E}{\hat{\chi}}+P_E-S_E\right)
+(1-\delta)p_0\left\{
 R_E-P_E
 +\frac{T_E-S_E}{\hat{\chi}}
 +\frac{\mu(T_E-R_E)-\eta(P_E-S_E)}{\mu-\eta} \right\},
\end{equation}
\begin{equation}\label{cond_kappa_ErrorDiscount3}
\kappa \le
R_E -\frac{\eta}{\mu-\eta}
\left(
 \frac{T_E-S_E}{\hat{\chi}} + T_E-R_E
\right)
-(1-\delta)(1-p_0)\left\{
 R_E-P_E
 +\frac{T_E-S_E}{\hat{\chi}}
 +\frac{\mu(P_E-S_E)-\eta(T_E-R_E)}{\mu-\eta}
\right\},
\end{equation}
\begin{equation}\label{cond_kappa_ErrorDiscount4}
\kappa\ge P_E+\frac{(\frac{1}{\delta}-1)p_0+\eta}{\mu-\eta}
\left(\frac{T_E-S_E}{\hat{\chi}}+P_E-S_E\right),
\end{equation}
(Appendix~\ref{ap:kappa}). The condition under which a positive $\kappa$ value that satisfies Eqs.~\eqref{cond_kappa_ErrorDiscount1}--\eqref{cond_kappa_ErrorDiscount4} exists is given by
\begin{equation}
\label{p0neq0_cond1}
\begin{split}
&P_E+\frac{\eta}{\mu-\eta}\left(\frac{T_E-S_E}{\hat{\chi}}+P_E-S_E\right)
+(1-\delta)p_0\left\{
 R_E-P_E
 +\frac{T_E-S_E}{\hat{\chi}}
 +\frac{\mu(T_E-R_E)-\eta(P_E-S_E)}{\mu-\eta}
\right\}\\
\le&R_E-\frac{(\frac{1}{\delta}-1)(1-p_0)+\eta}{\mu-\eta}\left(\frac{T_E-S_E}{\hat{\chi}}+T_E-R_E\right),
\end{split}
\end{equation}
\begin{equation}\label{p0neq0_cond2}
\begin{split}
&P_E+\frac{\eta}{\mu-\eta}\left(\frac{T_E-S_E}{\hat{\chi}}+P_E-S_E\right)
+(1-\delta)p_0\left\{
R_E-P_E
+\frac{T_E-S_E}{\hat{\chi}}
+\frac{\mu(T_E-R_E)-\eta(P_E-S_E)}{\mu-\eta}
\right\}\\
\le
&R_E
-\frac{\eta}{\mu-\eta}
\left(
 \frac{T_E-S_E}{\hat{\chi}}+T_E-R_E
\right)
-(1-\delta)(1-p_0)\left\{
 R_E-P_E
 +\frac{T_E-S_E}{\hat{\chi}}
 +\frac{\mu(P_E-S_E)-\eta(T_E-R_E)}{\mu-\eta}
\right\},
\end{split}
\end{equation}
\begin{equation}\label{p0neq0_cond3}
P_E+\frac{(\frac{1}{\delta}-1)p_0+\eta}{\mu-\eta}
\left(\frac{T_E-S_E}{\hat{\chi}}+P_E-S_E\right)
\le R_E-\frac{(\frac{1}{\delta}-1)(1-p_0)+\eta}{\mu-\eta}\left(\frac{T_E-S_E}{\hat{\chi}}+T_E-R_E\right),
\end{equation}
\begin{equation}\label{p0neq0_cond4}
\begin{split}
&P_E+\frac{(\frac{1}{\delta}-1)p_0+\eta}{\mu-\eta}\left(\frac{T_E-S_E}{\hat{\chi}}+P_E-S_E\right)\\
\le
&R_E
-\frac{\eta}{\mu-\eta}
\left(
 \frac{T_E-S_E}{\hat{\chi}}+T_E-R_E
\right)
-(1-\delta)(1-p_0)\left\{
 R_E-P_E
 +\frac{T_E-S_E}{\hat{\chi}}
 +\frac{\mu(P_E-S_E)-\eta(T_E-R_E)}{\mu-\eta}
\right\}.
\end{split}
\end{equation}
In this case, the above inequalities satisfy Eq.~\eqref{cond_kappa_errorNoDiscount1}. From the above inequalities,
\begin{eqnarray}
(1-\delta+2\delta\eta)(T_E-S_E)
\le\{\delta(\mu(R_E-P_E)-\eta(T_E-S_E))-(1-\delta)(T_E-R_E)\}\hat{\chi}\label{cond_chihat1},\\
(1-\delta+2\delta\eta)(T_E-S_E)\nonumber
\le \{\delta(\mu(R_E-P_E)-\eta(T_E-S_E))-(1-\delta)(P_E-S_E)\\
    -(1-\delta)p_0(T_E-R_E-P_E+S_E)\}\hat{\chi},\label{cond_chihat2}\\
(1-\delta+2\delta\eta)(T_E-S_E)\nonumber
\le \{\delta(\mu(R_E-P_E)-\eta(T_E-S_E))-(1-\delta)(T_E-R_E)\\
     +(1-\delta)p_0(T_E-R_E-P_E+S_E)\}\hat{\chi},\label{cond_chihat3}\\
(1-\delta+2\delta\eta)(T_E-S_E)
\le \{\delta(\mu(R_E-P_E)-\eta(T_E-S_E))-(1-\delta)(P_E-S_E)\}\hat{\chi},\label{cond_chihat4}
\end{eqnarray}
are obtained (Appendix~\ref{ap:chi}). Because the left-hand side of these inequalities are always greater than zero and $\hat{\chi}>0$, the coefficient of $\hat{\chi}$ must be greater than zero. Therefore, $\delta > \delta_c$ and $\eta < \eta_c$ must hold (Appendix~\ref{ap:deltac_etac}).
%When $R_E+P_E<T_E+S_E$, the first element of Eq.~\eqref{delta_c} and Eq.~\eqref{eta_c} is larger than the second one. 
%When $R_E+P_E>T_E+S_E$, the second element of Eq.~\eqref{delta_c} and Eq.~\eqref{eta_c} is  larger than the first one. 
%When $R_E+P_E=T_E+S_E$, both elements of Eq.~\eqref{delta_c} and Eq.~\eqref{eta_c} take the same value. 
Equations \eqref{cond_chihat1}--\eqref{cond_chihat4} are transformed into
\begin{equation}\label{chi_hat1}
\frac{(1-\delta+2\delta\eta)(T_E-S_E)}
{\delta(\mu(R_E-P_E)-\eta(T_E-S_E))-(1-\delta)(T_E-R_E)}
\le\hat{\chi},
\end{equation}
\begin{equation}\label{chi_hat2}
\frac{(1-\delta+2\delta\eta)(T_E-S_E)}
     {\delta(\mu(R_E-P_E)-\eta(T_E-S_E))-(1-\delta)(P_E-S_E)-(1-\delta)p_0(T_E-R_E-P_E+S_E)}
     \le\hat{\chi},
\end{equation}
\begin{equation}\label{chi_hat3}
\frac{(1-\delta+2\delta\eta)(T_E-S_E)}
{\delta(\mu(R_E-P_E)-\eta(T_E-S_E))
 -(1-\delta)(T_E-R_E)
 +(1-\delta)p_0(T_E-R_E-P_E+S_E)}
\le\hat{\chi},
\end{equation}
\begin{equation}\label{chi_hat4}
\frac{(1-\delta+2\delta\eta)(T_E-S_E)}
{\delta(\mu(R_E-P_E)-\eta(T_E-S_E))-(1-\delta)(P_E-S_E)}\le\hat{\chi}.
\end{equation}
When $R_E+P_E>T_E+S_E$, Eq.~\eqref{chi_hat4} is the maximum. When $R_E+P_E<T_E+S_E$, Eq.~\eqref{chi_hat1} is the maximum. When $R_E+P_E=T_E+S_E$,  Eqs.~\eqref{chi_hat1}--\eqref{chi_hat4} take the same value (Appendix~\ref{max_chi}). Therefore, when $0<p_0<1$, pcZD exists if and only if $\delta> \delta_c$ and $\eta<\eta_c$, where the $\delta_c$ and $\eta_c$ values coincide with those for Equalizer which are given by Eq.~\eqref{delta_c} and Eq.~\eqref{eta_c}. Under $\delta> \delta_c$ and $\eta<\eta_c$, Eqs.~\eqref{chi_hat1} and \eqref{chi_hat4} imply
\begin{equation}\label{eq:chi_c}
\chi\ge\chi_c\equiv \max(\chi_1,\chi_2),
\end{equation}
where
\begin{equation}
\chi_1=1+\frac{(1-\delta+2\delta\eta)(T_E-S_E)}
{\delta(\mu(R_E-P_E)-\eta(T_E-S_E))-(1-\delta)(T_E-R_E)},
\end{equation}
\begin{equation}
\chi_2=1+\frac{(1-\delta+2\delta\eta)(T_E-S_E)}
{\delta(\mu(R_E-P_E)-\eta(T_E-S_E))-(1-\delta)(P_E-S_E)}.
\end{equation}

{\bf (2) Case of $p_0=0$:}\\
In the case of $p_0=0$, Eqs.~\eqref{eq:Inequality1}--\eqref{eq:Inequality4} lead to
\begin{equation}
\hat{\chi}\left\{R_E-\kappa-\frac{\eta}{\mu-\eta}\left(\frac{T_E-S_E}{\hat{\chi}}+T_E-R_E\right) \right\}
\le\frac{1}{\phi}\le
\frac{\hat{\chi}\left\{R_E-\kappa-\frac{\eta}{\mu-\eta}\left(\frac{T_E-S_E}{\hat{\chi}}+T_E-R_E\right) \right\}}{1-\delta},
\end{equation}
\begin{equation}
\hat{\chi}\left\{R_E-\kappa+\frac{\mu}{\mu-\eta}\left(\frac{T_E-S_E}{\hat{\chi}}+T_E-R_E\right)\right\}
\le\frac{1}{\phi}
\le \frac{\hat{\chi}\left\{R_E-\kappa+\frac{\mu}{\mu-\eta}\left(\frac{T_E-S_E}{\hat{\chi}}+T_E-R_E\right)\right\}}{1-\delta},
\end{equation}
\begin{equation}
\frac{\hat{\chi}\left\{\kappa-P_E+\frac{\mu}{\mu-\eta}\left(\frac{T_E-S_E}{\hat{\chi}}+P_E-S_E\right)\right\}}{\delta} 
\le\frac{1}{\phi},
\end{equation}
\begin{equation}
\frac{\hat{\chi}\left\{\kappa-P_E-\frac{\eta}{\mu-\eta}\left(\frac{T_E-S_E}{\hat{\chi}}+P_E-S_E\right)\right\}}{\delta}
\le\frac{1}{\phi}.
\end{equation}
The condition under which a positive $\phi$ value that satisfies Eqs.~\eqref{cond_p00kappa1}--\eqref{cond_p00kappa2} exists is given by
\begin{equation}\label{cond_p00kappa1}
\hat{\chi}\left\{R_E-\kappa+\frac{\mu}{\mu-\eta}\left(\frac{T_E-S_E}{\hat{\chi}}+T_E-R_E\right)\right\}\le
\frac{\hat{\chi}\left\{R_E-\kappa-\frac{\eta}{\mu-\eta}\left(\frac{T_E-S_E}{\hat{\chi}}+T_E-R_E\right) \right\}}{1-\delta},
\end{equation}
\begin{equation}\label{cond_p00kappa2}
\frac{\hat{\chi}\left\{\kappa-P_E+\frac{\mu}{\mu-\eta}\left(\frac{T_E-S_E}{\hat{\chi}}+P_E-S_E\right)\right\}}{\delta} \le 
\frac{\hat{\chi}\left\{R_E-\kappa-\frac{\eta}{\mu-\eta}\left(\frac{T_E-S_E}{\hat{\chi}}+T_E-R_E\right) \right\}}{1-\delta}.
\end{equation}
Solving Eqs.~\eqref{cond_p00kappa1}--\eqref{cond_p00kappa2} for $\kappa$,
\begin{equation}
\kappa\le
R_E-\frac{(\frac{1}{\delta}-1)+\eta}{\mu-\eta}\left(\frac{T_E-S_E}{\hat{\chi}}+T_E-R_E\right),
\end{equation}
\begin{equation}
\kappa \le R_E
-\frac{\eta}{\mu-\eta}\left(\frac{T_E-S_E}{\hat{\chi}}+T_E-R_E\right)
-(1-\delta)\left\{
 R_E-P_E+
 \frac{T_E-S_E}{\hat{\chi}}
 +\frac{\mu(P_E-S_E)-\eta(T_E-R_E)}{\mu-\eta}
\right\}
\end{equation}
are obtained. If these inequalities hold, the second inequality of Eq.~\eqref{cond_kappa_errorNoDiscount1} is automatically satisfied. 
Thus, we only need to consider the two conditions above and the following condition:
\begin{equation}
P_E+\frac{\eta}{\mu-\eta}\left(\frac{T_E-S_E}{\hat{\chi}}+P_E-S_E\right)
\le\kappa.
\end{equation}
These three conditions are the same as the inequalities from Eq.~\eqref{cond_kappa_ErrorDiscount1} to Eq.~\eqref{cond_kappa_ErrorDiscount4} in the case of $p_0=0$. We obtain $\delta_c$, $\eta_c$ and $\chi_c$ as the case of $0<p_0<1$. Therefore, in the case of $p_0=0$, $\eta_c$ is given by Eq.~\eqref{eta_c}, $\delta_c$ is given by Eq.~\eqref{delta_c} and $\chi_c$ is given by Eq.~\eqref{eq:chi_c}.

{\bf (3) Case of $p_0=1$:}\\
In the case of $p_0=1$, Eqs.~\eqref{eq:Inequality1}--\eqref{eq:Inequality4} lead to
\begin{equation}
\frac{\hat{\chi}\left\{R_E-\kappa-\frac{\eta}{\mu-\eta}\left(\frac{T_E-S_E}{\hat{\chi}}+T_E-R_E\right) \right\}}{\delta}
\le\frac{1}{\phi},
\end{equation}
\begin{equation}
\frac{\hat{\chi}\left\{R_E-\kappa+\frac{\mu}{\mu-\eta}\left(\frac{T_E-S_E}{\hat{\chi}}+T_E-R_E\right)\right\}}{\delta}
\le\frac{1}{\phi},
\end{equation}
\begin{equation}
\hat{\chi}\left\{\kappa-P_E+\frac{\mu}{\mu-\eta}\left(\frac{T_E-S_E}{\hat{\chi}}+P_E-S_E\right)\right\}
\le\frac{1}{\phi}\le
\frac{\hat{\chi}\left\{\kappa-P_E+\frac{\mu}{\mu-\eta}\left(\frac{T_E-S_E}{\hat{\chi}}+P_E-S_E\right)\right\}}{1-\delta},
\end{equation}
\begin{equation}
\hat{\chi}\left\{\kappa-P_E-\frac{\eta}{\mu-\eta}\left(\frac{T_E-S_E}{\hat{\chi}}+P_E-S_E\right)\right\}
\le\frac{1}{\phi}\le
\frac{\hat{\chi}\left\{\kappa-P_E-\frac{\eta}{\mu-\eta}\left(\frac{T_E-S_E}{\hat{\chi}}+P_E-S_E\right)\right\}}{1-\delta}.
\end{equation}
The condition under which a positive $\phi$ value that satisfies Eqs.~\eqref{cond_p01kappa1} and \eqref{cond_p01kappa2} exists is given by
\begin{equation}\label{cond_p01kappa1}
\frac{\hat{\chi}\left\{R_E-\kappa+\frac{\mu}{\mu-\eta}\left(\frac{T_E-S_E}{\hat{\chi}}+T_E-R_E\right)\right\}}{\delta}\le
\frac{\hat{\chi}\left\{\kappa-P_E-\frac{\eta}{\mu-\eta}\left(\frac{T_E-S_E}{\hat{\chi}}+P_E-S_E\right)\right\}}{1-\delta},
\end{equation}
\begin{equation}\label{cond_p01kappa2}
\hat{\chi}\left\{\kappa-P_E+\frac{\mu}{\mu-\eta}\left(\frac{T_E-S_E}{\hat{\chi}}+P_E-S_E\right)\right\}\le 
\frac{\hat{\chi}\left\{\kappa-P_E-\frac{\eta}{\mu-\eta}\left(\frac{T_E-S_E}{\hat{\chi}}+P_E-S_E\right)\right\}}{1-\delta}.
\end{equation}
Solving Eqs.~\eqref{cond_p01kappa1} and \eqref{cond_p01kappa2} for $\kappa$,
\begin{equation}
\begin{split}
P_E+\frac{\eta}{\mu-\eta}\left(\frac{T_E-S_E}{\hat{\chi}}+P_E-S_E\right)
+(1-\delta)\left\{
 R_E-P_E
 +\frac{T_E-S_E}{\hat{\chi}}
 +\frac{\mu(T_E-R_E)-\eta(P_E-S_E)}{\mu-\eta}
\right\}
\le\kappa
\end{split},
\end{equation}
\begin{equation}
P_E
+\frac{\frac{1}{\delta}-1+\eta}{\mu-\eta}\left(\frac{T_E-S_E}{\hat{\chi}}+P_E-S_E\right)
\le \kappa.
\end{equation}
 If these inequalities hold, the second inequality of Eq.~\eqref{cond_kappa_errorNoDiscount1} is automatically satisfied. 
Thus, we only need to consider the two conditions above and the following condition:
\begin{equation}
\kappa \le R_E-\frac{\eta}{\mu-\eta}\left(\frac{T_E-S_E}{\hat{\chi}}+T_E-R_E\right).
\end{equation}
These three conditions are the same as the inequalities from Eq.~\eqref{cond_kappa_ErrorDiscount1} to Eq.~\eqref{cond_kappa_ErrorDiscount4} in the case of $p_0=1$. We obtain $\delta_c$, $\eta_c$ and $\chi_c$ as the case of $0<p_0<1$. Therefore, in the case of $p_0=1$, $\eta_c$ is given by Eq.~\eqref{eta_c}, $\delta_c$ is given by Eq.~\eqref{delta_c} and $\chi_c$ is given by Eq.~\eqref{eq:chi_c}. 

From the above, we derive the condition of $\eta$, $\delta$, $\kappa$ and $\chi$ where pcZD strategies exist. We reveal that the condition of $\delta$ is $\delta>\delta_c$, the condition of $\eta$ is given by $\eta<\eta_c$, the condition for $\kappa$ is given by Eq.~\eqref{cond_kappa_errorNoDiscount1} and the condition of $\chi$ is given by Eq.~\eqref{eq:chi_c}. The $\delta_c$ and $\eta_c$ are given by Eq.~\eqref{delta_c} and Eq.~\eqref{eta_c}. 
%\azumi{We found that the more $\delta$ is decreased, $\chi_c$ is increased. On the other hand, the range of $\kappa$ is not affected by discount factor $\delta$.}

$\chi_c$ with $\eta=0$ corresponds to Eq.~(67) and Eq.~(84) of Ref.~\cite{IchinoseMasuda2018JTheorBiol}. Ichinose and Masuda derived that $\chi_c$ takes the same value in both $\kappa=P$ and $\kappa=R$ cases \cite{IchinoseMasuda2018JTheorBiol}. Here we show that, not just $\kappa=P$ and $\kappa=R$, $\chi_c$ takes the same value for all $\kappa$ such that $P\le \kappa \le R$ when there are no observation errors but a discount factor ($\eta=0$ and $\delta<1$).

When there are observation errors and no discount factor ($\eta>0$ and $\delta=1$), Hao et al.~have numerically showed the conditions for $\kappa$ and $\chi$ \cite{Hao2015PhysRevE}. In our study, we analytically show the conditions for $\kappa$ and $\chi$, where $\kappa$ is given by Eq.~\eqref{cond_kappa_errorNoDiscount1} and $\chi$ is given by Eq.~\eqref{eq:chi_c} when $\delta=1$. 

%\subsection{Numerical examples}

\subsubsection{Numerical examples}
\begin{figure}[h]
  \centering
  \includegraphics[width=1\columnwidth]{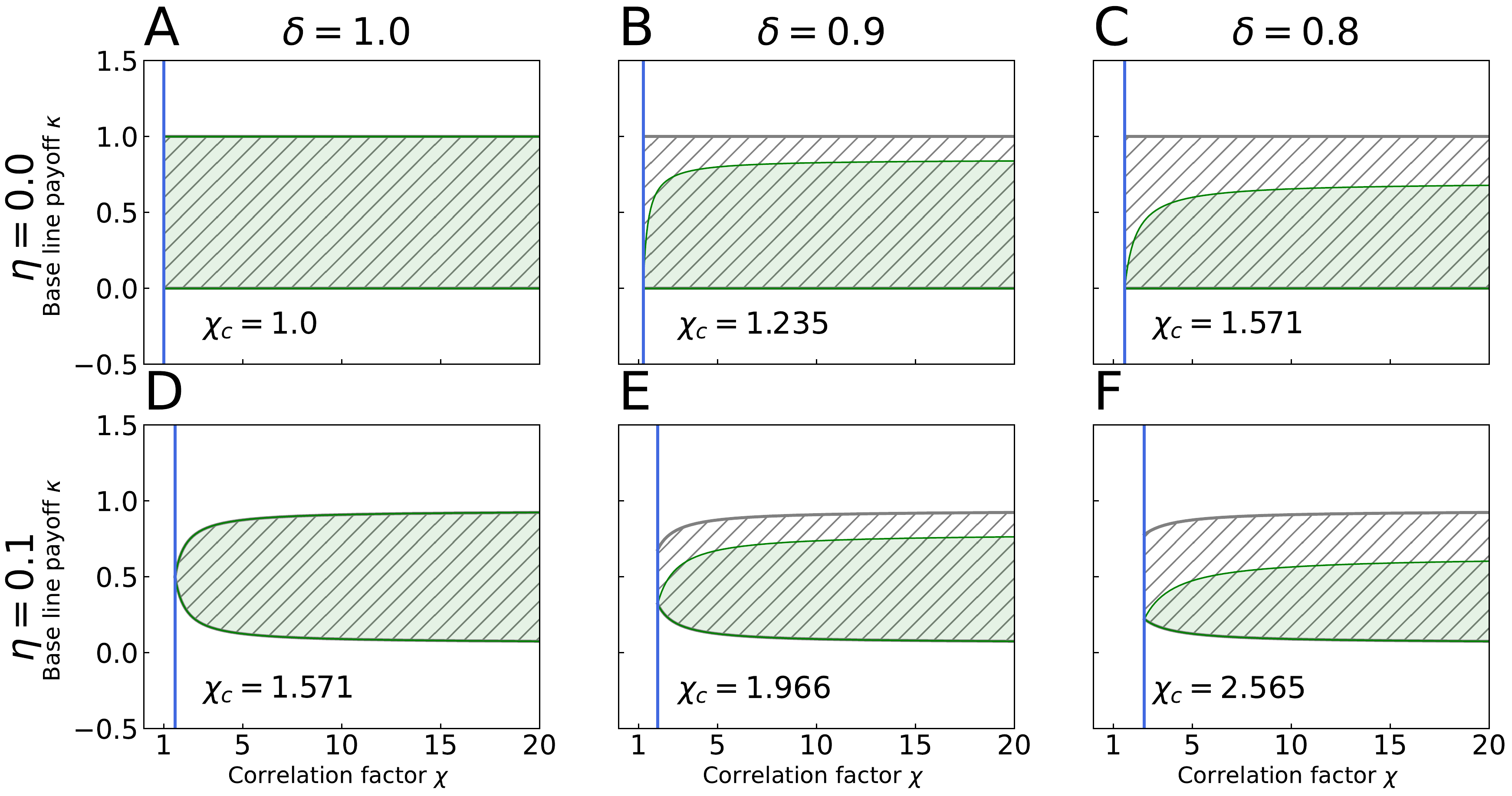}
  \caption{Possible regions of base line payoff $\kappa$ and correlation factor $\chi$ which pcZD can enforce. Green regions represent the possible $\kappa$ and $\chi$ for pcZD when $p_0=0$. Hatched regions, which include green regions as their subsets, represent the possible  $\kappa$ and $\chi$ for pcZD when $p_0$ is arbitrary.
Blue lines represent the minimum correlation factor $\chi_c$.
We set $(T_E, R_E, S_E, P_E)=(1.5, 0.5, 0.0, -0.5)$.
Each panel represents Eqs.~\eqref{cond_kappa_ErrorDiscount1}--\eqref{cond_kappa_ErrorDiscount4} and Eq.~\eqref{eq:chi_c} when error rates are $\eta=0.0,0.1$ and discount factors $\delta=1.0, 0.9, 0.8$, respectively.
When $\delta=1.0$, $\kappa$ and $\chi$ do not depend on $p_0$ because there is no discount factor.}
  \label{figure1}
\end{figure}

Figure \ref{figure1} shows the base line payoff $\kappa$ and correlation factor $\chi$ which pcZD can enforce when a discount factor and error rates change.
The figure represents Eqs.~\eqref{cond_kappa_ErrorDiscount1}--\eqref{cond_kappa_ErrorDiscount4} and Eq.~\eqref{eq:chi_c} when the error rates are $\eta=0.0,0.1$ and the discount factor $\delta=1.0, 0.9, 0.8$.
$T, R, P$, and $S$ are set so that $(T_E, R_E, S_E, P_E)=(1.5, 0.5, 0.0, -0.5)$ are satisfied.
Thus, the values of $T, R, P$ and $S$ change depending on the value of $\eta$.
The green regions represent the possible pcZD strategies when $p_0=0$.
When $p_0$ is arbitrary, the hatched regions including the green {regions} are possible.

Figures \ref{figure1}(A--C) are the cases of $\delta=1.0,0.9$ and $0.8$ when the error rate is fixed at zero ($\eta=0.0$).
As seen in the figures, $\chi_c$ becomes larger when the discount factor $\delta$ decreases.
On the other hand, the existing region of $\kappa$ which pcZD can enforce does not change when $p_0$ is arbitrary.
In Figs.~\ref{figure1}(B and C), when $p_0=0$, only $\kappa$ with $\kappa=P_E$ can take $\chi = \chi_c$.
This strategy corresponds to Extortion.
$p_0$ must be properly set when $\kappa \neq P_E$ is required.
For instance, $p_0$ must be $p_0=1$ when $\chi=\chi_c$ and $\kappa = R_E$ are required. This strategy corresponds to Generous.

Figures \ref{figure1}(D--F) are the cases of $\delta=1.0,0.9$, and $0.8$ when the error rate is fixed at 0.1 ($\eta=0.1$).
As well as the case of no errors, $\chi_c$ becomes larger when the discount factor $\delta$ decreases.
From these figures, we can see the region of $\kappa$ becomes small depending on $\eta$.
When $\delta=1.0$, only $\kappa$ with $\kappa=0.5$ can set $\chi=\chi_c$.
This fact has been numerically shown in Ref.~\cite{Hao2015PhysRevE}.
In contrast, in the cases of $\delta < 1.0$, the regions of $\kappa$ expand $\chi=\chi_c$, however, the regions are small compared to the case of $\eta=0$.
Moreover, $\chi_c$ becomes larger depending on $\eta$ and $\delta$.
Thus, the regions of $\chi$ and $\kappa$ become small even though the regions of $\kappa$ expand which can take $\chi=\chi_c$.

\begin{figure}[h]
  \centering
  \includegraphics[width=0.8\columnwidth]{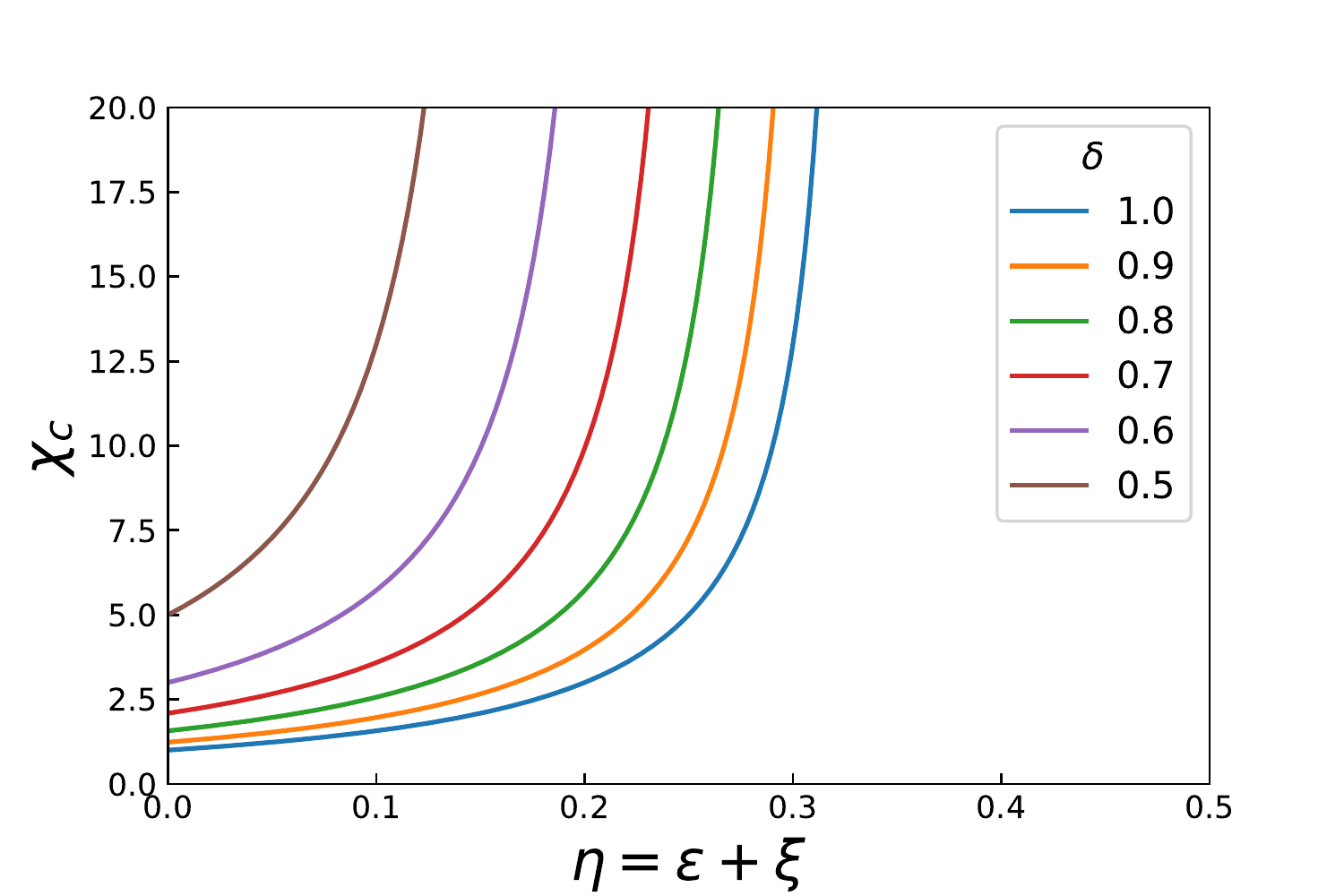}
  \caption{Minimum correlation factor $\chi_c$ for pcZD when the discount factor and the error rates are varied.}
  \label{figure3}
\end{figure}

Figure \ref{figure3} shows $\chi_c$ for pcZD when a discount factor and error rates are varied.
As seen in the figure, when there is neither a discount factor nor error rates, pcZD can take any $\chi$ with $\chi \geq 1$.
However, when there are errors or a discount factor, the range of $\chi$ becomes narrow.
As the error rate increases or the discount factor decreases from 1, the range of $\chi$ becomes narrower.
When $\delta=1$, $\chi_c$ are $1.0, 1.57143, 3.0$, and $13.0$ in $\eta =0,0.1,0.2,0.3$, respectively.
When $\delta=0.9$, $\chi_c$ are $1.2353, 1.9655, 3.9677$, and $33.0$ in $\eta =0,0.1,0.2,0.3$, respectively.
When $\delta=0.8$, $\chi_c$ are $1.5714, 2.5652$, and $5.7273$ in $\eta = 0,0.1,0.2$, respectively.
When $\delta=0.7$, $\chi_c$ are $2.0909, 3.5882$, and $9.9231$ in $\eta = 0,0.1,0.2$, respectively.
When $\delta=0.6$, $\chi_c$ are $3.0000, 5.7273$, and $33.0$ in $\eta = 0,0.1,0.2$, respectively.
When $\delta=0.5$, $\chi_c$ are $5.0$ and $13.0$ in $\eta =0,0.1$, respectively.
When $\chi$ is large, if the opponent improves his payoff, the payoff of the pcZD player is improved, more than the opponent.
Thus, for the opponent, the motivation that improves his payoff becomes weak because the range of improvement decreases.

\section{Conclusions}
In this study, we introduced observation errors and a discount factor in the RPD games and analytically investigated the conditions in which Equalizer or pcZD strategies can exist.
As a result, we obtained the conditions for the discount factor and the error rate where those strategies can exist.
As the error rates increase, the payoff of the opponent which Equalizer can control becomes narrower.
Moreover, as the error rates increase, Equalizer can exist only when a discount factor is high.
On the other hand, the ranges of the slope $\chi$ and the baseline payoff $\kappa$ for pcZD becomes narrow as a discount factor becomes small or the error rate becomes large.
In short, our results show that it is difficult for Equalizer and pcZD to exist when those factors are considered and that the controllability of the linear lines for Equalizer and pcZD decrease due to those two factors.

We have only considered observation errors but various other types of errors occur in animal and human behaviors.
The effect of such errors on cooperation has been studied by using the PD game \cite{Stephens1995JTheorBiol, Szolnoki2009PhysRevE, Nakamura2013PLoSComputBiol, Szolnoki2014NewJPhys}.
Implementation errors in the RPD game \cite{Stephens1995JTheorBiol}, noise for strategy updating in the spatial PD game \cite{Szolnoki2009PhysRevE}, incomplete observation in indirect reciprocity \cite{Nakamura2013PLoSComputBiol}, and deceitful defectors \cite{Szolnoki2014NewJPhys} have been incorporated.
It is important to explore the existence of ZD strategies under not just observation errors but also various types of errors.

It is well known that pcZD can change the adaptive opponent's strategy so that the opponent's payoff is improved.
As the error rates increase, the minimum $\chi$ becomes larger as we showed.
This means that pcZD exploits the opponent more, which decreases the motivation of the opponent to change his strategy because the improvement of the opponent's payoff becomes small.
If there are errors, because $\kappa$ can only take a value in $P_E < \kappa < R_E$, neither Extortion nor Generous can exist, which also means that it is impossible for pcZD to enforce an exploitative or a generous linear payoff relationship.
Chen and Zinger showed the existence of adapting paths which lead to unconditional cooperation for the adaptive opponent in the RPD game with $T>R>P>S$ and $2R>T+S>2P$ in the case of no errors \cite{ChenZinger2014JTheorBiol}.
It would be interesting to show the existence of such adapting paths even in the case with errors and a discount factor.

ZD strategies can unilaterally control the payoff of the opponent.
However, as we have shown, it would be difficult for those strategies to exist in the real world because real interactions are often noisy.
Although some previous studies have shown the role of ZD strategies in humans, one has to be careful how noises take place in such a situation.
Finally, our results are limited to the two-player RPD games. 
Other studies have focused on $n$-player games \cite{Hilbe2015JTheorBiol, Hilbe2014PNAS-zd, Pan2015SciRep-zd, GovaertCao2019arXiv}.
Investigating the conditions for the existence of ZD strategies in $n$-player games warrant future work.

\appendix
\setcounter{figure}{0}
\renewcommand{\thefigure}{\Alph{section}.\arabic{figure}}

\section{Any Equalizers do not exist when $\eta_c\le \eta<1/2$}\label{noLongerExist_Equalizer}
We prove that any Equalizers do not exist when $\eta_c\le \eta<1/2$.
First, when $\eta_c=\eta$ ($\Leftrightarrow \mu(R_E-P_E)-\eta(T_E-S_E)=0$), Eq.~\eqref{eq:equalizer_p2-p3} is not defined. Next, we consider the case of  $\eta_c < \eta<1/2$ ($\eta_c< \eta \Leftrightarrow \mu(R_E-P_E)-\eta(T_E-S_E) < 0$).
%We assume $\eta_c\le \eta<1/2$. 
In terms of Eq.~\eqref{eq:equalizer_p2-p3}, $0\le p_2\le 1$ and $0\le p_3\le 1$ lead to
\begin{equation}\label{eq:inequality_p3_lessthan0}
\mu(R_E-P_E)-\eta(T_E-S_E) \le 
p_1(\mu(T_E-P_E)-\eta(R_E-S_E))-(\frac{1}{\delta}+p_4)(T_E-R_E)
\le 0,
\end{equation}
\begin{equation}\label{eq:inequality_p4_lessthan0}
\mu(R_E-P_E)-\eta(T_E-S_E) \le 
(\frac{1}{\delta}-p_1)(P_E-S_E)+p_4(\mu(R_E-S_E)-\eta(T_E-P_E))
\le 0.
\end{equation}

We divide it into two cases (1) $\mu(T_E-P_E)-\eta(R_E-S_E)>0$ or $\mu(R_E-S_E)-\eta(T_E-P_E)>0$ and (2) $\mu(T_E-P_E)-\eta(R_E-S_E)\le0$ and $\mu(R_E-S_E)-\eta(T_E-P_E)\le0$.

First, we assume $\mu(T_E-P_E)-\eta(R_E-S_E)>0$ or $\mu(R_E-S_E)-\eta(T_E-P_E)>0$. From $\mu(T_E-P_E)-\eta(R_E-S_E)>0$, the first inequality of Eq.~\eqref{eq:inequality_p3_lessthan0} is not satisfied for any $p_1$ and $p_4$. Also, from $\mu(R_E-S_E)-\eta(T_E-P_E)>0$, the second inequality of Eq.~\eqref{eq:inequality_p4_lessthan0} is not satisfied for any $p_1$ and $p_4$.

Second, we assume $\mu(T_E-P_E)-\eta(R_E-S_E)\le0$ and $\mu(R_E-S_E)-\eta(T_E-P_E)\le0$. When $T_E-P_E>R_E-S_E$, $\mu(T_E-P_E)-\eta(R_E-S_E)\le0$ lead to 
\begin{equation}
\eta\ge\frac{T_E-P_E}{T_E+R_E-P_E-S_E}>1/2.
\end{equation}
Therefore, $\eta<1/2$ is not satisfied. When $T_E-P_E\le R_E-S_E$, $\mu(R_E-S_E)-\eta(T_E-P_E)\le0$ lead to
\begin{equation}
\eta\ge\frac{R_E-S_E}{T_E+R_E-P_E-S_E}\ge 1/2.
\end{equation}
Therefore, $\eta<1/2$ is not satisfied. As a result, any Equalizers do not exist when $\eta_c\le \eta<1/2$.

\section{ZD strategies with $0<\chi<1$ do not exist.}
\label{ap:chilessthan0}
Let us consider Eqs.~\eqref{eq:Inequality1}--\eqref{eq:Inequality4} under $\phi<0$ and $\chi<1$. In this case, we prove that ZD strategies with $0<\chi<1$ do not exist. 

We assume that $\phi<0$ and $\chi<1$. From here, we analyze them by dividing them into three cases (1) $0<p_0<1$, (2) $p_0=0$, (3) $p_0=1$ as follows.

{\bf (1) $0<p_0<1$:}\\
From Eqs.~\eqref{eq:Inequality1}--\eqref{eq:Inequality4},
\begin{eqnarray}
\frac{\hat{\chi}\left\{R_E-\kappa-\frac{\eta}{\mu-\eta}\left(\frac{T_E-S_E}{\hat{\chi}}+T_E-R_E\right) \right\}}{(1-\delta)(1-p_0)}
\le\frac{1}{\phi}\le
\frac{\hat{\chi}\left\{R_E-\kappa-\frac{\eta}{\mu-\eta}\left(\frac{T_E-S_E}{\hat{\chi}}+T_E-R_E\right) \right\}}{(1-\delta)(1-p_0)+\delta},
\\
\frac{\hat{\chi}\left\{R_E-\kappa+\frac{\mu}{\mu-\eta}\left(\frac{T_E-S_E}{\hat{\chi}}+T_E-R_E\right)\right\}}{(1-\delta)(1-p_0)}
\le\frac{1}{\phi}\le 
\frac{\hat{\chi}\left\{R_E-\kappa+\frac{\mu}{\mu-\eta}\left(\frac{T_E-S_E}{\hat{\chi}}+T_E-R_E\right)\right\}}{(1-\delta)(1-p_0)+\delta},
\\
\frac{\hat{\chi}\left\{\kappa-P_E+\frac{\mu}{\mu-\eta}\left(\frac{T_E-S_E}{\hat{\chi}}+P_E-S_E\right)\right\}}{(1-\delta)p_0}
\le \frac{1}{\phi}\le
\frac{\hat{\chi}\left\{\kappa-P_E+\frac{\mu}{\mu-\eta}\left(\frac{T_E-S_E}{\hat{\chi}}+P_E-S_E\right)\right\}}{(1-\delta)p_0+\delta} ,
\\
\frac{\hat{\chi}\left\{\kappa-P_E-\frac{\eta}{\mu-\eta}\left(\frac{T_E-S_E}{\hat{\chi}}+P_E-S_E\right)\right\}}{(1-\delta)p_0}
\le\frac{1}{\phi}\le
\frac{\hat{\chi}\left\{\kappa-P_E-\frac{\eta}{\mu-\eta}\left(\frac{T_E-S_E}{\hat{\chi}}+P_E-S_E\right)\right\}}{(1-\delta)p_0+\delta},
\end{eqnarray}
are obtained. For the above four equations, we need to check all sixteen ways of inequalities for $\phi$ to exist. As one of the conditions in those equations, the following inequality must hold
\begin{equation}
\frac{\hat{\chi}\left\{R_E-\kappa+\frac{\mu}{\mu-\eta}\left(\frac{T_E-S_E}{\hat{\chi}}+T_E-R_E\right)\right\}}{(1-\delta)(1-p_0)}
\le
\frac{\hat{\chi}\left\{R_E-\kappa-\frac{\eta}{\mu-\eta}\left(\frac{T_E-S_E}{\hat{\chi}}+T_E-R_E\right) \right\}}{(1-\delta)(1-p_0)+\delta}.
\end{equation}
From the above, we obtain
\begin{equation}
\begin{split}
\hat{\chi}\le
-\frac{((1-\delta)(1-p_0)+\delta\mu)(T_E-S_E)}
{
((1-\delta)(1-p_0)+\delta\mu)(T_E-S_E)
-((1-\delta)(1-p_0)+\delta\mu)(R_E-S_E)
+\delta(\mu-\eta)(R_E-\kappa)
}<-1,
\end{split}
\end{equation}
where the denominator is positive and $-((1-\delta)(1-p_0)+\delta\mu)(R_E-S_E)+\delta(\mu-\eta)(R_E-\kappa)$ is negative when $P_E \le \kappa \le R_E$. Therefore, ZD strategies with $0<\chi<1$ do not exist in the case of $0<p_0<1$ because we obtain $\chi<0$.

{\bf (2) $p_0=0$:}\\
From Eqs.~\eqref{eq:Inequality1}--\eqref{eq:Inequality4},
\begin{equation}
\frac{\hat{\chi}\left\{R_E-\kappa-\frac{\eta}{\mu-\eta}\left(\frac{T_E-S_E}{\hat{\chi}}+T_E-R_E\right) \right\}}{1-\delta}
\le\frac{1}{\phi}\le
\hat{\chi}\left\{R_E-\kappa-\frac{\eta}{\mu-\eta}\left(\frac{T_E-S_E}{\hat{\chi}}+T_E-R_E\right) \right\},
\end{equation}
\begin{equation}
\frac{\hat{\chi}\left\{R_E-\kappa+\frac{\mu}{\mu-\eta}\left(\frac{T_E-S_E}{\hat{\chi}}+T_E-R_E\right)\right\}}{1-\delta}
\le\frac{1}{\phi}\le
\hat{\chi}\left\{R_E-\kappa+\frac{\mu}{\mu-\eta}\left(\frac{T_E-S_E}{\hat{\chi}}+T_E-R_E\right)\right\},
\end{equation}
\begin{equation}
\frac{1}{\phi}\le
\frac{\hat{\chi}\left\{\kappa-P_E+\frac{\mu}{\mu-\eta}\left(\frac{T_E-S_E}{\hat{\chi}}+P_E-S_E\right)\right\}}{\delta},
\end{equation}
\begin{equation}
\frac{1}{\phi}\le
\frac{\hat{\chi}\left\{\kappa-P_E-\frac{\eta}{\mu-\eta}\left(\frac{T_E-S_E}{\hat{\chi}}+P_E-S_E\right)\right\}}{\delta},
\end{equation}
are obtained.
For the above four equations, we need to check all eight ways of inequalities for $\phi$ to exist. As one of the conditions in those equations, the following inequality must hold
\begin{equation}
\frac{\hat{\chi}\left\{R_E-\kappa+\frac{\mu}{\mu-\eta}\left(\frac{T_E-S_E}{\hat{\chi}}+T_E-R_E\right)\right\}}{1-\delta}
\le
\hat{\chi}\left\{R_E-\kappa-\frac{\eta}{\mu-\eta}\left(\frac{T_E-S_E}{\hat{\chi}}+T_E-R_E\right) \right\}.
\end{equation}
From the above, we obtain
\begin{equation}
\begin{split}
\hat{\chi} \le 
-\frac{(1-\delta\eta)(T_E-S_E)}
{(1-\delta\eta)(T_E-S_E)
-(1-\delta\eta)(R_E-S_E)
+\delta(\mu-\eta)(R_E-\kappa)}<-1,
\end{split}
\end{equation}
where the denominator is positive and $-(1-\delta\eta)(R_E-S_E)+\delta(\mu-\eta)(R_E-\kappa)$ is negative when $P_E \le \kappa \le R_E$. Therefore, ZD strategies with $0<\chi<1$ do not exist in the case of $p_0=0$ because we obtain $\chi<0$.

{\bf (3) $p_0=1$:}\\
From Eqs.~\eqref{eq:Inequality1}--\eqref{eq:Inequality4},
\begin{equation}
\frac{1}{\phi}\le
\frac{\hat{\chi}\left\{R_E-\kappa-\frac{\eta}{\mu-\eta}\left(\frac{T_E-S_E}{\hat{\chi}}+T_E-R_E\right) \right\}}{\delta},
\end{equation}
\begin{equation}
\frac{1}{\phi}\le 
\frac{\hat{\chi}\left\{R_E-\kappa+\frac{\mu}{\mu-\eta}\left(\frac{T_E-S_E}{\hat{\chi}}+T_E-R_E\right)\right\}}{\delta},
\end{equation}
\begin{equation}
\frac{\hat{\chi}\left\{\kappa-P_E+\frac{\mu}{\mu-\eta}\left(\frac{T_E-S_E}{\hat{\chi}}+P_E-S_E\right)\right\}}{1-\delta}
\le\frac{1}{\phi}\le
\hat{\chi}\left\{\kappa-P_E+\frac{\mu}{\mu-\eta}\left(\frac{T_E-S_E}{\hat{\chi}}+P_E-S_E\right)\right\},
\end{equation}
\begin{equation}
\frac{\hat{\chi}\left\{\kappa-P_E-\frac{\eta}{\mu-\eta}\left(\frac{T_E-S_E}{\hat{\chi}}+P_E-S_E\right)\right\}}{1-\delta}
\le\frac{1}{\phi}\le
\hat{\chi}\left\{\kappa-P_E-\frac{\eta}{\mu-\eta}\left(\frac{T_E-S_E}{\hat{\chi}}+P_E-S_E\right)\right\},
\end{equation}
are obtained. For the above four equations, we need to check all eight ways of inequalities for $\phi$ to exist. As one of the conditions in those equations, the following inequality must hold
\begin{equation}
\frac{\hat{\chi}\left\{\kappa-P_E+\frac{\mu}{\mu-\eta}\left(\frac{T_E-S_E}{\hat{\chi}}+P_E-S_E\right)\right\}}{1-\delta}
\le
\hat{\chi}\left\{\kappa-P_E-\frac{\eta}{\mu-\eta}\left(\frac{T_E-S_E}{\hat{\chi}}+P_E-S_E\right)\right\}.
\end{equation}
From the above, we obtain
\begin{equation}
\begin{split}
\hat{\chi}
\le
-\frac{(1-\delta\eta)(T_E-S_E)}
{(1-\delta\eta)(T_E-S_E)
-(1-\delta\eta)(T_E-P_E)
+\delta(\mu-\eta)(\kappa-P_E)}<-1,
\end{split}
\end{equation}
where the denominator is positive and $-(1-\delta\eta)(T_E-P_E)+\delta(\mu-\eta)(\kappa-P_E)$ is negative when $P_E \le \kappa \le R_E$. Therefore, ZD strategies with $0<\chi<1$ do not exist in the case of $p_0=1$ because we obtain $\chi<0$.

\section{Transforming Eqs.~\eqref{phi_cond1}--\eqref{phi_cond4} to Eqs.~\eqref{cond_kappa_ErrorDiscount1}--\eqref{cond_kappa_ErrorDiscount4}}\label{ap:kappa}
\subsection{Transforming Eq.~\eqref{phi_cond1} to Eq.~\eqref{cond_kappa_ErrorDiscount1}}
Multiplying both sides of Eq.~\eqref{phi_cond1} by 1/$\hat{\chi}$, $(1-\delta)(1-p_0)+\delta$ and $(1-\delta)(1-p_0)$, we obtain
\begin{equation}
\begin{split}
&(1-\delta)(1-p_0)\left\{R_E-\kappa+\frac{\mu}{\mu-\eta}\left(\frac{T_E-S_E}{\hat{\chi}}+T_E-R_E\right)\right\}
\\ \le
&((1-\delta)(1-p_0)+\delta)\left\{R_E-\kappa-\frac{\eta}{\mu-\eta}\left(\frac{T_E-S_E}{\hat{\chi}}+T_E-R_E\right)\right\}.
\end{split}
\end{equation}
Transposing the terms of $\kappa$ to the left-hand side and transposing the other term to the right-hand side, we obtain
\begin{equation}
\begin{split}
&((1-\delta)(1-p_0)+\delta)\kappa-(1-\delta)(1-p_0)\kappa
\\ \le
&((1-\delta)(1-p_0)+\delta)\left\{R_E-\frac{\eta}{\mu-\eta}\left(\frac{T_E-S_E}{\hat{\chi}}+T_E-R_E\right)\right\}\\
&-(1-\delta)(1-p_0)\left\{R_E+\frac{\mu}{\mu-\eta}\left(\frac{T_E-S_E}{\hat{\chi}}+T_E-R_E\right)\right\}.
\end{split}
\end{equation}
The above inequality can be simplified into
\begin{equation}
\delta\kappa
\le
\delta R_E-(\eta\delta+(1-\delta)(1-p_0))\frac{1}{\mu-\eta}\left(\frac{T_E-S_E}{\hat{\chi}}+T_E-R_E\right).
\end{equation}
Dividing both sides of the above inequality by $\delta$, we obtain
\begin{equation}
\kappa \le R_E-\frac{(\frac{1}{\delta}-1)(1-p_0)+\eta}{\mu-\eta}\left(\frac{T_E-S_E}{\hat{\chi}}+T_E-R_E\right).
\end{equation}

\subsection{Transforming Eq.~\eqref{phi_cond2} to Eq.~\eqref{cond_kappa_ErrorDiscount2}}
Multiplying both sides of Eq.~\eqref{phi_cond2} by 1/$\hat{\chi}$, $(1-\delta)(1-p_0)+\delta$ and $(1-\delta)p_0$, we obtain
\begin{equation}
\begin{split}
&(1-\delta)p_0\left\{R_E-\kappa+\frac{\mu}{\mu-\eta}\left(\frac{T_E-S_E}{\hat{\chi}}+T_E-R_E\right)\right\}
\\ \le
&((1-\delta)(1-p_0)+\delta)\left\{\kappa-P_E-\frac{\eta}{\mu-\eta}\left(\frac{T_E-S_E}{\hat{\chi}}+P_E-S_E\right)\right\}.
\end{split}
\end{equation}
Transposing the terms of $\kappa$ to the right-hand side and transposing the other term to the left-hand side, we obtain
\begin{equation}
\begin{split}
&(1-\delta)p_0\left\{R_E+\frac{\mu}{\mu-\eta}\left(\frac{T_E-S_E}{\hat{\chi}}+T_E-R_E\right)\right\}\\
&+((1-\delta)(1-p_0)+\delta)\left\{P_E+\frac{\eta}{\mu-\eta}\left(\frac{T_E-S_E}{\hat{\chi}}+P_E-S_E\right)\right\}
\\ \le
&((1-\delta)(1-p_0)+\delta)\kappa+(1-\delta)p_0\kappa.
\end{split}
\end{equation}
The above inequality can be simplified into
\begin{equation}
P_E
+\frac{\eta}{\mu-\eta}\left(\frac{T_E-S_E}{\hat{\chi}}+P_E-S_E\right)
+(1-\delta)p_0\left\{
R_E-P_E
+\frac{T_E-S_E}{\hat{\chi}}
+\frac{\mu(T_E-R_E)-\eta(P_E-S_E)}{\mu-\eta}
\right\}
\le \kappa.
\end{equation}

\subsection{Transforming Eq.~\eqref{phi_cond3} to Eq.~\eqref{cond_kappa_ErrorDiscount3}}
Multiplying both sides of Eq.~\eqref{phi_cond3} by 1/$\hat{\chi}$, $(1-\delta)p_0+\delta$ and $(1-\delta)(1-p_0)$, we obtain
\begin{equation}
\begin{split}
(1-\delta)(1-p_0)\left\{\kappa-P_E+\frac{\mu}{\mu-\eta}\left(\frac{T_E-S_E}{\hat{\chi}}+P_E-S_E\right)\right\}
\\ \le
((1-\delta)p_0+\delta)\left\{R_E-\kappa-\frac{\eta}{\mu-\eta}\left(\frac{T_E-S_E}{\hat{\chi}}+T_E-R_E\right)\right\}.
\end{split}
\end{equation}
Transposing the terms of $\kappa$ to the left-hand side and transposing the other term to the right-hand side, we obtain
\begin{equation}
\begin{split}
(1-\delta)(1-p_0)\kappa+((1-\delta)p_0+\delta)\kappa
\\ \le
((1-\delta)p_0+\delta)\left\{R_E-\frac{\eta}{\mu-\eta}\left(\frac{T_E-S_E}{\hat{\chi}}+T_E-R_E\right)\right\}\\
+(1-\delta)(1-p_0)\left\{P_E-\frac{\mu}{\mu-\eta}\left(\frac{T_E-S_E}{\hat{\chi}}+P_E-S_E\right)\right\}.
\end{split}
\end{equation}
The above inequality can be simplified into
\begin{equation}
\begin{split}
\kappa \le
R_E-\frac{\eta}{\mu-\eta}
\left(
 \frac{T_E-S_E}{\hat{\chi}}+T_E-R_E
\right)\\
-(1-\delta)(1-p_0)
\left\{
 R_E-P_E
 +\frac{T_E-S_E}{\hat{\chi}}
 +\frac{\mu(P_E-S_E)-\eta(T_E-R_E)}{\mu-\eta}
\right\}.
\end{split}
\end{equation}

\subsection{Transforming Eq.~\eqref{phi_cond4} to Eq.~\eqref{cond_kappa_ErrorDiscount4}}
Multiplying both sides of Eq.~\eqref{phi_cond4} by 1/$\hat{\chi}$, $(1-\delta)p_0+\delta$ and $(1-\delta)p_0$, we obtain
\begin{equation}
\begin{split}
(1-\delta)p_0\left\{\kappa-P_E+\frac{\mu}{\mu-\eta}\left(\frac{T_E-S_E}{\hat{\chi}}+P_E-S_E\right)\right\}
\\ \le
((1-\delta)p_0+\delta)\left\{\kappa-P_E-\frac{\eta}{\mu-\eta}\left(\frac{T_E-S_E}{\hat{\chi}}+P_E-S_E\right)\right\}.
\end{split}
\end{equation}
Transposing the terms of $\kappa$ to the right-hand side and transposing the other term to the left-hand side, we obtain
\begin{equation}
\begin{split}
(1-\delta)p_0\left\{-P_E+\frac{\mu}{\mu-\eta}\left(\frac{T_E-S_E}{\hat{\chi}}+P_E-S_E\right)\right\}\\
+((1-\delta)p_0+\delta)\left\{P_E+\frac{\eta}{\mu-\eta}\left(\frac{T_E-S_E}{\hat{\chi}}+P_E-S_E\right)\right\}\\
\le
((1-\delta)p_0+\delta)\kappa-(1-\delta)p_0\kappa.
\end{split}
\end{equation}
The above inequality can be simplified into
\begin{equation}
\delta P_E+((1-\delta)p_0+\delta\eta)\frac{1}{\mu-\eta}
\left(\frac{T_E-S_E}{\hat{\chi}}+P_E-S_E\right)\le\delta\kappa.
\end{equation}
Dividing both sides of the above inequality by $\delta$, we obtain
\begin{equation}
P_E+\frac{(\frac{1}{\delta}-1)p_0+\eta}{\mu-\eta}
\left(\frac{T_E-S_E}{\hat{\chi}}+P_E-S_E\right)\le\kappa.
\end{equation}

\section{Transforming Eqs.~\eqref{p0neq0_cond1}--\eqref{p0neq0_cond4} to Eqs.~\eqref{cond_chihat1}--\eqref{cond_chihat4}}\label{ap:chi}
\subsection{Transforming Eq.~\eqref{p0neq0_cond1} to Eq.~\eqref{cond_chihat1}}
Transposing the terms of $\hat{\chi}$ to left-hand side of Eq.~\eqref{p0neq0_cond1} and transposing the other term to right-hand side of Eq.~\eqref{p0neq0_cond1}, we obtain
\begin{equation}
\begin{split}
&\left\{\frac{\eta}{\mu-\eta}+(1-\delta)p_0+\frac{(\frac{1}{\delta}-1)(1-p_0)+\eta}{\mu-\eta}\right\}\frac{T_E-S_E}{\hat{\chi}}\\
\le 
& R_E-P_E-\frac{(\frac{1}{\delta}-1)(1-p_0)+\eta}{\mu-\eta}
\left(
 T_E-R_E
\right)\\
&-\frac{\eta}{\mu-\eta}\left(P_E-S_E\right)
-(1-\delta)p_0\left\{
 R_E-P_E
 +\frac{\mu(T_E-R_E)-\eta(P_E-S_E)}{\mu-\eta}
\right\}.
\end{split}
\end{equation}
Multiplying both sides of the above inequality by $(\mu-\eta)\delta$, we obtain
\begin{equation}
\begin{split}
&\{\delta \eta+(\mu-\eta)\delta(1-\delta)p_0+(1-\delta)(1-p_0)+\delta\eta\}\frac{T_E-S_E}{\hat{\chi}}\\
\le 
& (\mu-\eta)\delta(R_E-P_E)
 -\left\{(1-\delta)(1-p_0)+\delta\eta\right\}
\left(
 T_E-R_E
\right)\\
&-\eta\delta\left(P_E-S_E\right)
-\delta(1-\delta)p_0\left\{
 (\mu-\eta)(R_E-P_E)
 +\mu(T_E-R_E)-\eta(P_E-S_E)
\right\}.
\end{split}
\end{equation}
The above inequality can be simplified into
\begin{equation}
\begin{split}
&(1-(1-\delta)p_0)(1-\delta+2\delta\eta)\frac{T_E-S_E}{\hat{\chi}}\\
\le 
& (\mu-\eta)\delta(R_E-P_E)
 -(1-\delta)(1-p_0)(T_E-R_E)-\delta\eta (T_E-R_E)\\
&-\eta\delta\left(P_E-S_E\right)
 -\delta(1-\delta)p_0(\mu-\eta)(R_E-P_E)
 -\delta(1-\delta)p_0(\mu(T_E-R_E)-\eta(P_E-S_E)).
\end{split}
\end{equation}
Factoring out a common factor of the above inequality, we obtain
\begin{equation}
\begin{split}
&(1-(1-\delta)p_0)(1-\delta+2\delta\eta)\frac{T_E-S_E}{\hat{\chi}}\\
\le 
&(1-(1-\delta)p_0)\delta(\mu-\eta)(R_E-P_E)
 -(1-(1-\delta)p_0)\eta\delta(P_E-S_E)\\
&-((1-\delta)(1-p_0)+\delta(1-\delta)p_0\mu+\delta\eta)(T_E-R_E).
\end{split}
\end{equation}
The above inequality can be simplified into
\begin{equation}
\begin{split}
&(1-(1-\delta)p_0)(1-\delta+2\delta\eta)\frac{T_E-S_E}{\hat{\chi}}\\
\le 
&(1-(1-\delta)p_0)\delta(\mu-\eta)(R_E-P_E)
 -(1-(1-\delta)p_0)\eta\delta(P_E-S_E)\\
&-(1-\mu\delta)(1-(1-\delta)p_0)(T_E-R_E).
\end{split}
\end{equation}
Dividing both sides of the above inequality by $1-(1-\delta)p_0$, we obtain
\begin{equation}
(1-\delta+2\delta\eta)\frac{T_E-S_E}{\hat{\chi}}\\
\le \delta(\mu(R_E-P_E)-\eta(T_E-S_E))-(1-\delta)(T_E-R_E).
\end{equation}
Multiplying both sides of the above inequality by $\hat{\chi}$, we obtain
\begin{equation}
(1-\delta+2\delta\eta)(T_E-S_E)
\le (\delta(\mu(R_E-P_E)-\eta(T_E-S_E))-(1-\delta)(T_E-R_E))\hat{\chi}.
\end{equation}

\subsection{Transforming Eq.~\eqref{p0neq0_cond2} to Eq.~\eqref{cond_chihat2}}
Transposing the terms of $\hat{\chi}$ to left-hand side of Eq.~\eqref{p0neq0_cond2} and transposing the other term to right-hand side of Eq.~\eqref{p0neq0_cond2}, we obtain
\begin{equation}
\begin{split}
&\left\{
 \frac{\eta}{\mu-\eta}
 +(1-\delta)p_0
 +\frac{\eta}{\mu-\eta}+(1-\delta)(1-p_0)
\right\}\frac{T_E-S_E}{\hat{\chi}}\\
\le
&R_E-P_E
 -\frac{\eta}{\mu-\eta}(T_E-R_E)
 -(1-\delta)(1-p_0)(R_E-P_E)
 -(1-\delta)(1-p_0)\frac{\mu(P_E-S_E)-\eta(T_E-R_E)}{\mu-\eta}\\
&-\frac{\eta}{\mu-\eta}(P_E-S_E)-(1-\delta)p_0(R_E-P_E)-(1-\delta)p_0\frac{\mu(T_E-R_E)-\eta(P_E-S_E)}{\mu-\eta}.\\
\end{split}
\end{equation}
Multiplying both sides of the above inequality by $\mu-\eta$, we obtain
\begin{equation}
\begin{split}
&\left\{
 (\mu-\eta)(1-\delta)p_0
 +2\eta
 +(\mu-\eta)(1-\delta)(1-p_0)
\right\}\frac{T_E-S_E}{\hat{\chi}}\\
\le
&(\mu-\eta)(R_E-P_E)
 -\eta(T_E-R_E)
 -(\mu-\eta)(1-\delta)(1-p_0)(R_E-P_E)\\
&-(1-\delta)(1-p_0)(\mu(P_E-S_E)-\eta(T_E-R_E))\\
&-\eta(P_E-S_E)
 -(\mu-\eta)(1-\delta)p_0(R_E-P_E)
 -(1-\delta)p_0(\mu(T_E-R_E)-\eta(P_E-S_E)).\\
\end{split}
\end{equation}
The above inequality can be simplified into
\begin{equation}
\begin{split}
&(1-\delta+2\eta\delta)\frac{T_E-S_E}{\hat{\chi}}\\
\le
&(\mu-\eta)(R_E-P_E)
 -\eta(T_E-R_E)
 -(\mu-\eta)(1-\delta)(1-p_0)(R_E-P_E)\\
&-\mu(1-\delta)(1-p_0)(P_E-S_E)+\eta(1-\delta)(1-p_0)(T_E-R_E)\\
&-\eta(P_E-S_E)
 -(\mu-\eta)(1-\delta)p_0(R_E-P_E)
 -\mu(1-\delta)p_0(T_E-R_E)+\eta(1-\delta)p_0(P_E-S_E).\\
\end{split}
\end{equation}
Factoring out a common factor of the above inequality, we obtain
\begin{equation}
\begin{split}
(1-\delta+2\eta\delta)\frac{T_E-S_E}{\hat{\chi}}
\le
&(\mu-\eta)(1-(1-\delta)(1-p_0)-(1-\delta)p_0)(R_E-P_E)\\
&+(-\eta+\eta(1-\delta)(1-p_0)-\mu(1-\delta)p_0)(T_E-R_E)\\
&+(-\mu(1-\delta)(1-p_0)-\eta+\eta(1-\delta)p_0)(P_E-S_E).
\end{split}
\end{equation}
The above inequality can be simplified into
\begin{equation}
\begin{split}
(1-\delta+2\eta\delta)\frac{T_E-S_E}{\hat{\chi}}
\le
&\delta(\mu(R_E-P_E)-\eta(T_E-S_E))\\
&-(1-\delta)(P_E-S_E)-(1-\delta)p_0(T_E-R_E-P_E+S_E).
\end{split}
\end{equation}
Multiplying both sides of the above inequality by $\hat{\chi}$, we obtain
\begin{equation}
\begin{split}
(1-\delta+2\eta\delta)(T_E-S_E)
\le
&(\delta(\mu(R_E-P_E)-\eta(T_E-S_E))\\
&-(1-\delta)(P_E-S_E)-(1-\delta)p_0(T_E-R_E-P_E+S_E))\hat{\chi}.
\end{split}
\end{equation}

\subsection{Transforming Eq.~\eqref{p0neq0_cond3} to Eq.~\eqref{cond_chihat3}}
Transposing the terms of $\hat{\chi}$ to left-hand side of Eq.~\eqref{p0neq0_cond3} and transposing the other term to right-hand side of Eq.~\eqref{p0neq0_cond3}, we obtain
\begin{equation}
\begin{split}
&\left\{
 \frac{(\frac{1}{\delta}-1)p_0+\eta}{\mu-\eta}
 +\frac{(\frac{1}{\delta}-1)(1-p_0)+\eta}{\mu-\eta}
 \right\}\frac{T_E-S_E}{\hat{\chi}}\\
&\le R_E-P_E
 -\frac{(\frac{1}{\delta}-1)(1-p_0)+\eta}{\mu-\eta}(T_E-R_E)
 -\frac{(\frac{1}{\delta}-1)p_0+\eta}{\mu-\eta}(P_E-S_E).
\end{split}
\end{equation}
Multiplying both sides of the above inequality by $(\mu-\eta)\delta$, we obtain
\begin{equation}
\begin{split}
&\left(
 (1-\delta)p_0
 +\delta\eta
 +(1-\delta)(1-p_0)
 +\delta\eta
 \right)\frac{T_E-S_E}{\hat{\chi}}\\
&\le (\mu-\eta)\delta(R_E-P_E)
 -(\left(1-\delta)(1-p_0)+\delta\eta\right)(T_E-R_E)
 -(\left(1-\delta)p_0+\delta\eta\right)(P_E-S_E).
\end{split}
\end{equation}
The above inequality can be simplified into
\begin{equation}
\begin{split}
&\left(
 1-\delta
 +2\delta\eta
 \right)\frac{T_E-S_E}{\hat{\chi}}\\
&\le 
 \delta(\mu(R_E-P_E)-\eta(T_E-S_E))
 -(1-\delta)(T_E-R_E)
 +(1-\delta)p_0(T_E-R_E-P_E+S_E).
\end{split}
\end{equation}
Multiplying both sides of the above inequality by $\hat{\chi}$, we obtain
\begin{equation}
\begin{split}
&\left(
 1-\delta
 +2\delta\eta
 \right)(T_E-S_E)\\
&\le 
 (\delta(\mu(R_E-P_E)-\eta(T_E-S_E))
 -(1-\delta)(T_E-R_E)
 +(1-\delta)p_0(T_E-R_E-P_E+S_E))\hat{\chi}.
\end{split}
\end{equation}

\subsection{Transforming Eq.~\eqref{p0neq0_cond4} to Eq.~\eqref{cond_chihat4}}
Transposing the terms of $\hat{\chi}$ to left-hand side of Eq.~\eqref{p0neq0_cond4} and transposing the other term to right-hand side of Eq.~\eqref{p0neq0_cond4}, we obtain
\begin{equation}
\begin{split}
&\left\{\frac{(\frac{1}{\delta}-1)p_0+\eta}{\mu-\eta}+(1-\delta)(1-p_0)
+\frac{\eta}{\mu-\eta}\right\}\frac{T_E-S_E}{\hat{\chi}}\\
\le
&R_E-P_E
 -\frac{\eta}{\mu-\eta}(T_E-R_E)
 -(1-\delta)(1-p_0)(R_E-P_E)\\
&-(1-\delta)(1-p_0)\frac{\mu(P_E-S_E)-\eta(T_E-R_E)}{\mu-\eta}
-\frac{(\frac{1}{\delta}-1)p_0+\eta}{\mu-\eta}(P_E-S_E).
\end{split}
\end{equation}
Multiplying both sides of the above inequality by $(\mu-\eta)\delta$, we obtain
\begin{equation}
\begin{split}
&\left\{
(1-\delta)p_0
+\delta\eta
+\delta(\mu-\eta)(1-\delta)(1-p_0)
+\delta\eta
\right\}\frac{T_E-S_E}{\hat{\chi}}\\
\le
&(\mu-\eta)\delta(R_E-P_E)
 -\eta\delta(T_E-R_E)
 -(\mu-\eta)\delta(1-\delta)(1-p_0)(R_E-P_E)\\
&-(1-\delta)(1-p_0)\delta(\mu(P_E-S_E)-\eta(T_E-R_E))
-((1-\delta)p_0+\eta\delta)(P_E-S_E).
\end{split}
\end{equation}
Factoring out a common factor of the above inequality, we obtain
\begin{equation}
\begin{split}
&\left\{
(1-\delta)p_0
+\delta\eta
+\delta(\mu-\eta)(1-\delta)(1-p_0)
+\delta\eta
\right\}\frac{T_E-S_E}{\hat{\chi}}\\
\le
&(\mu-\eta)(1-(1-\delta)(1-p_0))\delta(R_E-P_E)+((1-\delta)(1-p_0)\eta-\eta)\delta(T_E-R_E)\\
&+(-(1-\delta)(1-p_0)\delta\mu-((1-\delta)p_0+\delta\eta))(P_E-S_E).
\end{split}
\end{equation}
The above inequality can be simplified into
\begin{equation}
\begin{split}
&(1-\delta+2\delta\eta)(1-(1-\delta)(1-p_0))\frac{T_E-S_E}{\hat{\chi}}\\
\le
&(\mu-\eta)\delta(1-(1-\delta)(1-p_0))(R_E-P_E)\\
&-\delta(1-(1-\delta)(1-p_0))\eta(T_E-R_E)+(\delta\mu-1)(1-(1-\delta)(1-p_0))(P_E-S_E).
\end{split}
\end{equation}
Dividing both sides of the above inequality by $1-(1-\delta)(1-p_0)$, we obtain
\begin{equation}
(1-\delta+2\delta\eta)\frac{T_E-S_E}{\hat{\chi}}
\le (\mu-\eta)\delta(R_E-P_E)-\delta\eta(T_E-R_E)+(\delta\mu-1)(P_E-S_E).
\end{equation}
The above inequality can be simplified into
\begin{equation}
(1-\delta+2\delta\eta)\frac{T_E-S_E}{\hat{\chi}}
\le \delta(\mu(R_E-P_E)-\eta(T_E-S_E))-(1-\delta)(P_E-S_E).
\end{equation}
Multiplying the both sides of the above inequality by $\hat{\chi}$, we obtain
\begin{equation}
(1-\delta+2\delta\eta)(T_E-S_E)
\le (\delta(\mu(R_E-P_E)-\eta(T_E-S_E))-(1-\delta)(P_E-S_E))\hat{\chi}.
\end{equation}

\section{Condition of discount factor $\delta$ and error rate $\eta$ for the existence of pcZD strategies}
\label{ap:deltac_etac}
Because the coefficient of $\hat{\chi}$ of Eqs.~\eqref{cond_chihat1}--\eqref{cond_chihat4} must be positive, we obtain 
\begin{equation}\label{delta_cond1}
\delta(\mu(R_E-P_E)-\eta(T_E-S_E))-(1-\delta)(T_E-R_E)>0,
\end{equation}
\begin{equation}\label{delta_cond2}
\delta(\mu(R_E-P_E)-\eta(T_E-S_E))-(1-\delta)(P_E-S_E+p_0(T_E-R_E-P_E+S_E))>0,
\end{equation}
\begin{equation}\label{delta_cond3}
 \delta(\mu(R_E-P_E)-\eta(T_E-S_E))-(1-\delta)(T_E-R_E-p_0(T_E-R_E-P_E+S_E))>0,
\end{equation}
\begin{equation}\label{delta_cond4}
\delta(\mu(R_E-P_E)-\eta(T_E-S_E))-(1-\delta)(P_E-S_E)>0.
\end{equation}
%Therefore, we obtain
%\begin{equation}
%\mu(R_E-P_E)-\eta(T_E-S_E)>0 \Leftrightarrow \eta<\frac{R_E-P_E}{T_E+R_E-P_E-S_E}.
%\end{equation}
In this case, solving Eqs.~\eqref{delta_cond1}--\eqref{delta_cond4} for $\delta$, we obtain
\begin{equation}\label{delta1}
 \delta > \frac{T_E-R_E}{\mu(R_E-P_E)-\eta(T_E-S_E)+T_E-R_E},
\end{equation}
\begin{equation}\label{delta2}
\delta > \frac{P_E-S_E+p_0(T_E-R_E-P_E+S_E)}{\mu(R_E-P_E)-\eta(T_E-S_E)+P_E-S_E+p_0(T_E-R_E-P_E+S_E)},
\end{equation}
\begin{equation}\label{delta3}
 \delta > 
 \frac{T_E-R_E-p_0(T_E-R_E-P_E+S_E)}
 {\mu(R_E-P_E)-\eta(T_E-S_E)+T_E-R_E-p_0(T_E-R_E-P_E+S_E)},
\end{equation}
\begin{equation}\label{delta4}
 \delta > \frac{P_E-S_E}{\mu(R_E-P_E)-\eta(T_E-S_E)+P_E-S_E}.
\end{equation}
For Eqs.~\eqref{delta1}--\eqref{delta4}, due to $\delta<1$, $\mu(R_E-P_E)-\eta(T_E-S_E)>0 (\Leftrightarrow \eta<\eta_c)$ must hold.
Thus, there are no pcZD strategies when $\eta>\eta_c$. $\eta_c$ is the same as the one defined inEq.~\eqref{eta_c}.
In the case of $R_E+P_E>T_E+S_E$, Eq.~\eqref{delta4} is the maximum. In the case of $R_E+P_E<T_E+S_E$, Eq.~\eqref{delta1} is the maximum. In the case of $R_E+P_E=T_E+S_E$, Eqs.~\eqref{delta1}--\eqref{delta4} take the same value (Appendix~\ref{ap:delta}). Therefore, we obtain $\delta > \delta_c$. $\delta_c$ is the same as the one defined in Eq.~\eqref{delta_c}.

\section{Magnitude relationship of Eqs.~\eqref{delta1}--\eqref{delta4}}
In this section, we check the magnitude relationship of Eqs.~\eqref{delta1}--\eqref{delta4}.
\label{ap:delta}

\subsection{Case of $R_E+P_E>T_E+S_E$}
We first check the case of $R_E+P_E>T_E+S_E$.
Each of the differences between Eq.~\eqref{delta4} and Eqs.~\eqref{delta1}--\eqref{delta3} is

\noindent
(Eq.~\eqref{delta4}$-$Eq.~\eqref{delta1})
\begin{equation}\label{eq:PS_TR}
\begin{split}
 &=\frac{P_E-S_E}{\mu(R_E-P_E)-\eta(T_E-S_E)+P_E-S_E}
   -\frac{T_E-R_E}{\mu(R_E-P_E)-\eta(T_E-S_E)+T_E-R_E}\\
 &=-\frac{(T_E-R_E-P_E+S_E)(\mu(R_E-P_E)-\eta(T_E-S_E))}
 {(\mu(R_E-P_E)-\eta(T_E-S_E)+P_E-S_E)(\mu(R_E-P_E)-\eta(T_E-S_E)+T_E-R_E)}>0.
\end{split}
\end{equation}
(Eq.~\eqref{delta4}$-$Eq.~\eqref{delta2})
\begin{equation}
\begin{split}
 =&\frac{P_E-S_E}{\mu(R_E-P_E)-\eta(T_E-S_E)+P_E-S_E}\\
   &-\frac{P_E-S_E+p_0(T_E-R_E-P_E+S_E)}
   {\mu(R_E-P_E)-\eta(T_E-S_E)+P_E-S_E+p_0(T_E-R_E-P_E+S_E)}\\
 =&\{(P_E-S_E)(\mu(R_E-P_E)-\eta(T_E-S_E)+P_E-S_E+p_0(T_E-R_E-P_E+S_E))\\
   &\quad -(\mu(R_E-P_E)-\eta(T_E-S_E)+P_E-S_E)(P_E-S_E+p_0(T_E-R_E-P_E+S_E))\}\\
  & /\{(\mu(R_E-P_E)-\eta(T_E-S_E)+P_E-S_E)(\mu(R_E-P_E)-\eta(T_E-S_E)+P_E-S_E+p_0(T_E-R_E-P_E+S_E))\}\\
 =&-\{p_0(T_E-R_E-P_E+S_E)(\mu(R_E-P_E)-\eta(T_E-S_E))\}\\
  & /\{(\mu(R_E-P_E)-\eta(T_E-S_E)+P_E-S_E)(\mu(R_E-P_E)-\eta(T_E-S_E)+P_E-S_E+p_0(T_E-R_E-P_E+S_E))\}\\
   &>0.
\end{split}
\end{equation}
(Eq.~\eqref{delta4}$-$Eq.~\eqref{delta3})
\begin{equation}
\begin{split}
 =&\frac{P_E-S_E}{\mu(R_E-P_E)-\eta(T_E-S_E)+P_E-S_E}\\
   &-\frac{T_E-R_E-p_0(T_E-R_E-P_E+S_E)}
 {\mu(R_E-P_E)-\eta(T_E-S_E)+(T_E-R_E)-p_0(T_E-R_E-P_E+S_E)}\\
 =&\{(P_E-S_E)(\mu(R_E-P_E)-\eta(T_E-S_E)+(T_E-R_E)-p_0(T_E-R_E-P_E+S_E))\\
  &\quad -(T_E-R_E-p_0(T_E-R_E-P_E+S_E))(\mu(R_E-P_E)-\eta(T_E-S_E)+P_E-S_E) \}\\
  &/\{(\mu(R_E-P_E)-\eta(T_E-S_E)+P_E-S_E)(\mu(R_E-P_E)-\eta(T_E-S_E)+(T_E-R_E)-p_0(T_E-R_E-P_E+S_E))\}\\
 =&-\{(1-p_0)(T_E-R_E-P_E+S_E)(\mu(R_E-P_E)-\eta(T_E-S_E))\}\\
  &/\{(\mu(R_E-P_E)-\eta(T_E-S_E)+P_E-S_E)(\mu(R_E-P_E)-\eta(T_E-S_E)+(T_E-R_E)-p_0(T_E-R_E-P_E+S_E))\}\\
 &>0.
\end{split}
\end{equation}
Therefore, in this case, Eq.~\eqref{delta4} is the maximum.

\subsection{Case of $R_E+P_E<T_E+S_E$}
Next, we check the case of $R_E+P_E<T_E+S_E$.
Each of the differences between Eq.~\eqref{delta1} and Eqs.~\eqref{delta2}--\eqref{delta4} is

\noindent
(Eq.~\eqref{delta1}$-$Eq.~\eqref{delta2})
\begin{equation}
\begin{split}
 =& \frac{T_E-R_E}{\mu(R_E-P_E)-\eta(T_E-S_E)+T_E-R_E}\\
  &-\frac{P_E-S_E+p_0(T_E-R_E-P_E+S_E)}
         {\mu(R_E-P_E)-\eta(T_E-S_E)+(P_E-S_E)+p_0(T_E-R_E-P_E+S_E)}\\
 =& \frac{(1-p_0)(T_E-R_E-P_E+S_E)(\mu(R_E-P_E)-\eta(T_E-S_E))}
         {(\mu(R_E-P_E)-\eta(T_E-S_E)+T_E-R_E)(\mu(R_E-P_E)-\eta(T_E-S_E)+(P_E-S_E)+p_0(T_E-R_E-P_E+S_E))}\\
         &>0.
\end{split}
\end{equation}
(Eq.~\eqref{delta1}$-$Eq.~\eqref{delta3})
\begin{equation}
\begin{split}
 =&\frac{T_E-R_E}{\mu(R_E-P_E)-\eta(T_E-S_E)+T_E-R_E}
 -\frac{T_E-R_E-p_0(T_E-R_E-P_E+S_E)}
 {\mu(R_E-P_E)-\eta(T_E-S_E)+(T_E-R_E)-p_0(T_E-R_E-P_E+S_E)}\\
 =&\{
      (T_E-R_E)(\mu(R_E-P_E)-\eta(T_E-S_E)+(T_E-R_E)-p_0(T_E-R_E-P_E+S_E))\\
      &-(T_E-R_E-p_0(T_E-R_E-P_E+S_E))(\mu(R_E-P_E)-\eta(T_E-S_E)+T_E-R_E)
   \}\\
     &/\{(\mu(R_E-P_E)-\eta(T_E-S_E)+T_E-R_E)(\mu(R_E-P_E)-\eta(T_E-S_E)+(T_E-R_E)-p_0(T_E-R_E-P_E+S_E))\}\\
 =&\frac{
         p_0(T_E-R_E-P_E+S_E)(\mu(R_E-P_E)-\eta(T_E-S_E))
         }
       {(\mu(R_E-P_E)-\eta(T_E-S_E)+T_E-R_E)(\mu(R_E-P_E)-\eta(T_E-S_E)+(T_E-R_E)-p_0(T_E-R_E-P_E+S_E))}\\
       &>0.
\end{split}
\end{equation}
(Eq.~\eqref{delta1}$-$Eq.~\eqref{delta4})
\begin{equation}
\begin{split}\label{eq:TR_PS}
 &=\frac{T_E-R_E}{\mu(R_E-P_E)-\eta(T_E-S_E)+T_E-R_E}-\frac{P_E-S_E}{\mu(R_E-P_E)-\eta(T_E-S_E)+P_E-S_E}\\
 &=\frac{(T_E-R_E)(\mu(R_E-P_E)-\eta(T_E-S_E)+P_E-S_E)
        -(P_E-S_E)(\mu(R_E-P_E)-\eta(T_E-S_E)+T_E-R_E)}
 {(\mu(R_E-P_E)-\eta(T_E-S_E)+P_E-S_E)(\mu(R_E-P_E)-\eta(T_E-S_E)+T_E-R_E)}\\
 &=\frac{(T_E-R_E-P_E+S_E)(\mu(R_E-P_E)-\eta(T_E-S_E))}
 {(\mu(R_E-P_E)-\eta(T_E-S_E)+P_E-S_E)(\mu(R_E-P_E)-\eta(T_E-S_E)+T_E-R_E)}>0.
\end{split}
\end{equation}
Therefore, in this case, Eq.~\eqref{delta1} is the maximum.

\subsection{Case of $R_E+P_E=T_E+S_E$}
In the case of $R_E+P_E=T_E+S_E$, Eqs.~\eqref{eq:PS_TR}--\eqref{eq:TR_PS} become zero. Therefore, Eqs.~\eqref{delta1}--\eqref{delta4} take the same value.

\section{Magnitude relationship of Eqs.~\eqref{chi_hat1}--\eqref{chi_hat4}}\label{max_chi}
In this section we check the magnitude relationship of Eqs.~\eqref{chi_hat1}--\eqref{chi_hat4}.

\subsection{Case of $R_E+P_E>T_E+S_E$}
We first check the case of $R_E+P_E>T_E+S_E$.
Each of the differences between Eq.~\eqref{chi_hat4} and Eqs.~\eqref{chi_hat1}--\eqref{chi_hat3} is

\noindent
(Denominator of Eq.~\eqref{chi_hat1})$-$(Denominator of Eq.~\eqref{chi_hat4})
\begin{equation}\label{eq:chi_hat1Minuschi_hat4}
\begin{split}
=&\delta(\mu(R_E-P_E)-\eta(T_E-S_E))-(1-\delta)(T_E-R_E)\\
 &-(\delta(\mu(R_E-P_E)-\eta(T_E-S_E))-(1-\delta)(P_E-S_E))\\
=&-(1-\delta)(T_E-R_E-P_E+S_E)>0.
\end{split}
\end{equation}
(Denominator of Eq.~\eqref{chi_hat2})$-$(Denominator of Eq.~\eqref{chi_hat4})
\begin{equation}\label{eq:chi_hat2Minuschi_hat4}
\begin{split}
=&\delta(\mu(R_E-P_E)-\eta(T_E-S_E))-(1-\delta)(P_E-S_E)-(1-\delta)p_0(T_E-R_E-P_E+S_E)\\
 &-(\delta(\mu(R_E-P_E)-\eta(T_E-S_E))-(1-\delta)(P_E-S_E))\\
=&-(1-\delta)p_0(T_E-R_E-P_E+S_E)>0.
\end{split}
\end{equation}
(Denominator of Eq.~\eqref{chi_hat3})$-$(Denominator of Eq.~\eqref{chi_hat4})
\begin{equation}\label{eq:chi_hat3Minuschi_hat4}
\begin{split}
=&\delta(\mu(R_E-P_E)-\eta(T_E-S_E))-(1-\delta)(T_E-R_E)+(1-\delta)p_0(T_E-R_E-P_E+S_E)\\
 &-(\delta(\mu(R_E-P_E)-\eta(T_E-S_E))-(1-\delta)(P_E-S_E))\\
=&-(1-\delta)(1-p_0)(T_E-R_E-P_E+S_E)>0.
\end{split}
\end{equation}
Therefore, in this case, Eq.~\eqref{chi_hat4} is the maximum.

\subsection{Case of $R_E+P_E<T_E+S_E$}
Next, we check the case of $R_E+P_E<T_E+S_E$. Each of the differences between Eq.~\eqref{chi_hat1} and Eqs.~\eqref{chi_hat2}--\eqref{chi_hat4} is

\noindent
(Denominator of Eq.~\eqref{chi_hat2})$-$(Denominator of Eq.~\eqref{chi_hat1})
\begin{equation}
\begin{split}
=&\delta(\mu(R_E-P_E)-\eta(T_E-S_E))-(1-\delta)(P_E-S_E)-(1-\delta)p_0(T_E-R_E-P_E+S_E)\\
 &-(\delta(\mu(R_E-P_E)-\eta(T_E-S_E))-(1-\delta)(T_E-R_E))\\
=&(1-\delta)(1-p_0)(T_E-R_E-P_E+S_E)>0.
\end{split}
\end{equation}
(Denominator of Eq.~\eqref{chi_hat3})$-$(Denominator of Eq.~\eqref{chi_hat1})
\begin{equation}
\begin{split}
=&\delta(\mu(R_E-P_E)-\eta(T_E-S_E))-(1-\delta)(T_E-R_E)+(1-\delta)p_0(T_E-R_E-P_E+S_E)\\
 &-(\delta(\mu(R_E-P_E)-\eta(T_E-S_E))-(1-\delta)(T_E-R_E))\\
=&(1-\delta)p_0(T_E-R_E-P_E+S_E)>0.
\end{split}
\end{equation}
(Denominator of Eq.~\eqref{chi_hat4})$-$(Denominator of Eq.~\eqref{chi_hat1})
\begin{equation}
\begin{split}
=&\delta(\mu(R_E-P_E)-\eta(T_E-S_E))-(1-\delta)(P_E-S_E)\\
 &-(\delta(\mu(R_E-P_E)-\eta(T_E-S_E))-(1-\delta)(T_E-R_E))\\
=&(1-\delta)(T_E-R_E-P_E+S_E)>0.
\end{split}
\end{equation}
Therefore, in this case, Eq.~\eqref{chi_hat1} is the maximum.

\subsection{Case of $R_E+P_E=T_E+S_E$}
In the case of $R_E+P_E=T_E+S_E$, Eqs.~\eqref{eq:chi_hat1Minuschi_hat4}--\eqref{eq:chi_hat3Minuschi_hat4} become zero. Therefore, Eqs.~\eqref{chi_hat1}--\eqref{chi_hat4} take the same value.

\section*{Acknowledgment}
This study was partly supported by JSPS KAKENHI Grant Number JP19K04903 (G.I.).

%\bibliographystyle{unsrt}
%\bibliography{citations-zd}

\end{document}